\documentstyle[12pt]{article}
\input xy
\xyoption{all}

\def\hilo{{}_{{}_{{}_{{}_{{}_{}}}}} {}^{{}^{{}^{}}}}
\newcommand{\bm}{\begin{matrix}}
\def\lrdd{\left ( ~}
\def\rrdd{\hilo \right )}
\newcommand{\up}[1]{^{[{#1}]}{}}
\newcommand{\kket}[1]{\left | {#1} \right \rangle }
\newcommand{\bbra}[1]{\left \langle {#1} \right | }
\def\d{\delta}

\def\hh{{1\over 2}}
\newcommand{\bbb}{\begin{displaymath}}
\newcommand{\eee}{\end{displaymath}}

\begin{document}

\hfill VPI-IPNAS-07-07 

\hfill UTTG-06-07

\vspace{0.1in}

\begin{center}

{\large\bf Non-Birational Twisted Derived Equivalences}

{\large\bf in Abelian GLSMs}

\vspace{0.1in}

Andrei C\u ald\u araru$^1$,
Jacques Distler$^2$, Simeon Hellerman$^3$, Tony Pantev$^4$,
Eric Sharpe$^5$ \\

$\,$

\begin{tabular}{cc}
{ \begin{tabular}{c}
$^1$ Mathematics Department\\
University of Wisconsin \\
Madison, WI  53706-1388\\
\end{tabular} } &
{\begin{tabular}{c}
$^2$ University of Texas, Austin\\
Department of Physics\\
Austin, TX  78712-0264
\end{tabular} } \\
{ \begin{tabular}{c}
$^3$ School of Natural Sciences \\
Institute for Advanced Study \\
Princeton, NJ  08540 
\end{tabular} } &
{ \begin{tabular}{c}
$^4$ Department of Mathematics\\
University of Pennsylvania \\
Philadelphia, PA  19104-6395 
\end{tabular} } \\
{ \begin{tabular}{c}
$^5$ Physics Department\\
Virginia Tech \\
Blacksburg, VA  24061
\end{tabular} }
& $\,$ 
\end{tabular}

{\tt andreic@math.wisc.edu}, {\tt distler@golem.ph.utexas.edu},
{\tt simeon@ias.edu}, {\tt tpantev@math.upenn.edu}, 
{\tt ersharpe@vt.edu}\\

$\,$

\end{center}

In this paper we discuss some examples of abelian gauged linear sigma models
realizing twisted derived equivalences between non-birational spaces,
and realizing geometries in novel fashions.
Examples of gauged linear sigma models with non-birational K\"ahler phases
are a relatively new phenomenon.
Most of our examples involve gauged linear sigma models for complete
intersections of quadric hypersurfaces, though we also discuss
some more general cases and their interpretation. 
We also propose a more general understanding of the relationship
between K\"ahler phases of gauged linear sigma models,
namely that they are related by (and realize) Kuznetsov's
`homological projective duality.'
Along the way, we shall see how `noncommutative spaces' (in Kontsevich's
sense) are realized physically in gauged linear sigma models, providing
examples of new types of conformal field theories.
Throughout, the physical realization of stacks plays a key role in
interpreting physical structures appearing in GLSMs, and we find that
stacks are implicitly much more common in GLSMs than previously realized.

\begin{flushleft}
September 2007
\end{flushleft}

\newpage

\tableofcontents

\newpage

\section{Introduction}

Gauged linear sigma models, first described in \cite{WitPhases}, have proven
to be a crucial tool for string compactifications.
They have provided insight into topics ranging from the structure
of SCFT moduli spaces to curve-counting in Calabi-Yau's.

When a GLSM describes different geometries in different limits of
K\"ahler moduli space,
it has long been assumed that the different geometries are birational to
one another, {\it e.g.} related by flops, blowups, blowdowns, or other
such transformations.  It has also been assumed that the only Calabi-Yau's
one could describe as phases of GLSM's 
were built as complete intersections in toric
varieties or flag manifolds (or other semiclassical moduli spaces of
supersymmetric gauge theories).
However, recently we have begun to learn that neither statement is
always the case. 

In \cite{ps4}[section 12.2] and then in \cite{hori1},
examples have been given of 
gauged linear sigma models involving
(a) a Calabi-Yau not presented as a complete intersection, and
(b) two non-birationally-equivalent Calabi-Yau's.
In \cite{hori1}, a nonabelian GLSM was analyzed,
describing a complete intersection in a Grassmannian was shown
to lie on the same moduli space as the vanishing locus of a Pfaffian,
and in \cite{ps4}[section 12.2] an abelian GLSM was analyzed,
describing complete intersection of four degree two hypersurfaces
in ${\bf P}^7$ at one limit and a branched double cover of
${\bf P}^3$, branched over a degree eight hypersurface 
(Clemens' octic double solid) in another K\"ahler phase.

In this paper, we shall study further examples of
abelian GLSM's describing non-birational K\"ahler phases.
We begin by working through the example of \cite{ps4}[section 12.2]
in much greater detail, then go on to consider other examples.
One natural question this work poses is:  is there a mathematical
relationship between the different K\"ahler phases, some notion
that replaces `birational'?  We propose that the different
K\"ahler geometric phases of a given GLSM should all be understood
as being related by `homological projective duality,' a recent concept
introduced into mathematics by Kuznetsov.
Put another way, we propose that GLSM's implicitly give a physical
realization of Kuznetsov's homological projective duality.

In addition, we argue that new kinds of conformal field theories are
realized as these duals.  These are physical realizations of
Kuznetsov's noncommutative resolutions
of singular spaces.
We introduce these new conformal field theories and discuss
some of their basic properties, but clearly a great deal of work 
should be done to properly understand them and their role in physics.

The analysis of the Landau-Ginzburg points of these GLSMs revolves
around subtleties in the two-dimensional abelian gauge theories
with nonminimal charges, which provide one physical realization
of strings on gerbes.  In other words, this paper describes in
detail one application of gerbes and stacks.  The original application
of the technology of stacks, aside from the completely obvious possibility
of enlarging the number of possible string compactifications,
was to understand physical properties of string orbifolds such as
the fact that they give well-behaved CFT's \cite{estx}.
More recent applications outlined in \cite{ps4} range from making
physical predictions for certain quantum cohomology computations
to reconciling different physical aspects of the geometric Langlands
program.

We begin in section~\ref{quadrics-section} with a detailed
analysis of the GLSM for ${\bf P}^7[2,2,2,2]$.
We find, after an analysis that involves understanding how stacks
appear physically, and also after finding a crucial Berry phase,
that the Landau-Ginzburg point seems, on the face of it,
to be in the same universality class as a nonlinear sigma model
on a branched double cover of ${\bf P}^3$, which is another
Calabi-Yau.  This is already interesting in that these two
geometries, the complete intersection and the branched double cover,
are {\it not} birational to one another, violating the conventional
wisdom that different geometric K\"ahler phases of the same GLSM
should be birational to one another.  This is also noteworthy for
the novel realization of the geometry at the Landau-Ginzburg point, 
as something other than the simultaneous vanishing locus of a set of
$F$-terms, realizing a complete intersection in a toric variety.
Further analysis reveals further subtleties:
although for analogues in lower dimensions the branched double cover
at the Landau-Ginzburg point is smooth, for the particular
example ${\bf P}^7[2,2,2,2]$ the branched double cover is 
mathematically singular, whereas the GLSM does not exhibit
any singularities.  An additional study 
leads us to believe that the structure
actually being realized is a `noncommutative resolution'
of the singular branched double cover, a conjecture which is verified
by studying matrix factorizations at the Landau-Ginzburg point.
(Noncommutative resolutions are defined by their sheaf theory,
so, seeing that matrix factorizations match the mathematics nails down
the interpretation as a noncommutative resolution.)
In particular, this means that we are getting some new conformal field
theories -- CFT's that look like ordinary nonlinear sigma models
on smooth patches, but which are fundamentally different over
singular parts of the classical geometry.
We tentatively identify this duality between the large-radius and
Landau-Ginzburg point geometries as an example of Kuznetsov's
`homological projective duality.'
Finally, at the end of section~\ref{quadrics-section} 
we also outline how this
generalizes in other dimensions.

In section~\ref{vwex} we discuss another Calabi-Yau example of
this phenomenon, in which a GLSM for a complete intersection of
quadrics has a (noncommutative resolution of a) branched double
cover at its Landau-Ginzburg point.  This particular example
amounts to a fibered version of a low-dimensional example of the
form from section~\ref{quadrics-section}, and is also closely related
to geometries appearing in Vafa and Witten's work on discrete torsion
\cite{vafawitten}.  We also discuss how deformation theory issues,
the last remaining property of discrete torsion that has not been
completely explicitly derived from $B$ fields, can be understood
from the perspective of noncommutative spaces.

In section~\ref{non-cy} we extend these considerations to a series
of non-Calabi-Yau examples, in which again we see GLSM's relating
complete intersections of quadrics to (noncommutative resolutions of)
branched double covers.

In section~\ref{genl-ci} we extend these notions to more general
complete intersections, not of quadrics.  We find that homological
projective duality continues to apply to more general cases,
even cases in which the Landau-Ginzburg point does not have a 
geometric or nc-geometric interpretation,
and we conjecture that all phases of all gauged linear sigma models 
are related by homological projective duality.

Finally in appendix~\ref{cy-categories} we review some general aspects
of noncommutative resolutions and nc spaces, to make this paper
more nearly self-contained, as these notions have
not, to our knowledge, been previously discussed in the physics
literature.

There are many technical similarities between the abelian GLSMs for
complete intersections described in \cite{ps4}[section 12.2]
and the nonabelian GLSMs describing complete intersections in
Grassmannians in \cite{hori1}[section 5].
In both cases, the geometry at one limit of the GLSM K\"ahler moduli
space is realized in a novel fashion:  here and in 
\cite{ps4}[section 12.2] as a double cover realized by gerbes
and a nonminimally-charged gauge theory,
in \cite{hori1}[section 5], through strong-coupling nonabelian gauge dynamics.
In both cases, the geometries at either end of the GLSM K\"ahler moduli
space are not birational, but instead are related by Kuznetsov's homological
projective duality.
In both cases the superpotential has the form
\begin{displaymath}
W(\Phi) \: = \:  \sum_{ij} \Phi_i A^{ij} \Phi_j
\end{displaymath}
for some matrix $A$, giving a mass to the chiral superfields $\Phi_i$.
The primary physical difference between the gauged linear sigma model
in \cite{hori1}[section 5] and \cite{ps4}[section 12.2]
is that in the
former, at least one $\phi$ always remains massless (and is removed by quantum
corrections), whereas in the latter all of the $\phi$ are generically
massive.  Thus, in the latter case one generically has a nonminimally
charged field, $p$, and so gerbes are relevant, whereas in the
former there is never a nonminimally-charged-field story.

In \cite{meron}, further nonabelian examples were presented,
expanding on that discussed in \cite{hori1}[section 5],
and the relevance of homological projective duality, discussed in
more detail in this paper, was introduced.

The physics of complete intersections of quadrics plays a central
role in this paper.
More information on the mathematics of complete intersections of
quadrics can be found in, for example, \cite{reidthesis}.

\section{Quadrics in projective space and branched double covers}
\label{quadrics-section}

Our first example involves a gauged linear sigma model describing
a complete intersection of four quadrics in ${\bf P}^7$ in the
$r \gg 0$ limit, and a double cover of ${\bf P}^3$ branched over
a degree $8$ locus in the $r \ll 0$ limit.
This example originally appeared in \cite{ps4}[section 12.2]; we shall review
and elaborate upon that example here.

\subsection{Review of the mathematics}  \label{rev-quadrics}

We shall begin by reviewing pertinent mathematics.
First, let us remind the reader why
a double cover of ${\bf P}^3$ branched over a degree 8
hypersurface in ${\bf P}^3$ is an example of a Calabi-Yau.

Let $B$ be a complex manifold and let $D \subset B$ be a smooth divisor.
A double cover $\pi: S \rightarrow B$ branched along $D$ is specified 
uniquely by a holomorphic line bundle $L \rightarrow B$,
such that $L^{\otimes 2} \cong {\cal O}_B(D)$.  Explicitly, if
$s \in H^0(B, L^{\otimes 2})$ is a section with divisor $D$,
then the double cover $S$ is the divisor in the total space of $L$
given by the equation $z^2 = p^* s$, where $p: \mbox{Tot}(L)
\rightarrow B$ is the natural projection, and
$z \in H^0(\mbox{Tot}(L), p^*L)$ is the tautological section.
For such a cover $S$ the adjunction formula gives
\begin{displaymath}
K_S \: = \: \pi^*(K_B \otimes L)
\end{displaymath}
In particular, if $L = K_B^{-1}$, then the double cover $S$ will have
a trivial canonical class.

In the present case, the base is ${\bf P}^3$, with canonical bundle
of degree $-4$, and so we see that the branched double cover is
Calabi-Yau if the branch locus has degree $8$.
For a closely related discussion in the context of a different
example, see \cite[chapter 4.4, p. 548]{gh}.
Double covers of ${\bf P}^3$ branched over a degree 8 hypersurface in
${\bf P}^3$ are known as octic double solids, and are described
in greater detail in {\it e.g.} \cite{clemens1,cynk1}.

Mathematically, the double cover can be understood
as a moduli space of certain bundles on the complete intersection
of quadrics.  (Each quadric in ${\bf P}^7$ carries two
distinct spinor bundles which restrict to bundles on the complete
intersection, and when the quadric degenerates, the spinor bundles
become isomorphic, hence giving the double cover of ${\bf P}^3$.)

Now, the twisted\footnote{Twisted in the sense described in
\cite{meshelandrei}:  because of a flat $B$ field present, transition
functions only close up to cocycles on triple overlaps.
See \cite{gp} for a discussion of the Brauer group of 
${\bf P}^7[2,2,2,2]$.} derived category of coherent sheaves of the
branched double cover of ${\bf P}^3$ has been expected
\cite{grosspriv}, and was recently proven\footnote{
What was proven in \cite{kuz2} was a relation between the twisted
derived category of a noncommutative resolution of the branched double cover,
and the derived category of ${\bf P}^7[2,2,2,2]$.
That noncommutative resolution will play an important role in the physics,
as we shall discuss later.
} \cite{kuz2}, to be
isomorphic to the derived category of a
complete intersection of four quadrics in ${\bf P}^7$.
Specifically, there is a twisted derived equivalence if the double
cover and the complete intersection are related as follows.
Let $Q_a$ denote the four quadrics in the complete intersection,
and consider the following linear combination:
\begin{displaymath}
\sum_a p_a Q_a(x)
\end{displaymath}
where the $p_a$ are homogeneous coordinates on ${\bf P}^3$.
Rewrite this linear combination as
\begin{displaymath}
\sum_{ij} x_i A^{ij}(p) x_j
\end{displaymath}
where $A^{ij}$ is an $8 \times 8$ matrix with entries linear in the
$p_a$.
Then the complete intersection of the four quadrics $Q_a$ is
twisted derived equivalence to a branched double cover of ${\bf P}^3$
branched over the degree eight locus $\mbox{det }A = 0$.

Such derived equivalences are not unusual in gauged linear sigma models.
After all, derived categories encapsulate the open string B model
\cite{medc,dougdc,medcrev}, and the B model is independent of K\"ahler
moduli, hence one expects that different geometries on the same GLSM
K\"ahler moduli space will have isomorphic derived categories.

On the other hand, it is also typically the case that
different phases of a GLSM will be related by birational transformations,
and that is not the case here:
as pointed out by M.~Gross
\cite{grosspriv} the complete intersection in
${\bf P}^7$ has no contractible curves, whereas the branched double
cover has several ordinary double points.

\subsection{Basic GLSM analysis}

In this section we will work through the analysis of a gauged linear
sigma model describing
the complete intersection
of four degree-two hypersurfaces in ${\bf P}^7$ at large radius.
We will find, after careful analysis involving an understanding of
how gerbes appear in physics, that the Landau-Ginzburg point of this
GLSM can be interpreted geometrically as a branched double cover
of ${\bf P}^3$, the same branched double cover related to the
complete intersection by a twisted derived equivalence.

This gauged linear sigma model has a total of twelve chiral superfields,
eight ($\phi_i$, $i \in \{ 1, \cdots, 8 \}$) of charge 1
corresponding to
homogeneous coordinates on ${\bf P}^7$, and four ($p_a$,
$a \in \{ 1, \cdots, 4 \}$) of charge $-2$ corresponding
to the four hypersurfaces.

The D-term for this gauged linear sigma model reads
\begin{displaymath}
\sum_i | \phi_i |^2 \: - \: 2 \sum_a | p_a |^2 \: = \: r
\end{displaymath}

When $r \gg 0$, then we see that not all the $\phi_i$ can vanish,
corresponding to their interpretation as homogeneous coordinates
on ${\bf P}^7$.  More generally, for $r \gg 0$ we recover the
geometric interpretation of this gauged linear sigma model as
a complete intersection of quadrics.

For $r \ll 0$, we find a different story.
There, the D-term constraint says that not all the $p_a$'s can
vanish; in fact, the $p_a$'s act as homogeneous coordinates on
a ${\bf P}^3$, except that these homogeneous coordinates have
charge 2 rather than charge 1.  

Because of those nonminimal charges, the Landau-Ginzburg point is
ultimately going to describe a (branched) double cover.
The superpotential
\begin{displaymath}
W \: = \: \sum_i p_i Q_i(\phi)
\end{displaymath}
(where the $Q_i$ are quadric polynomials)
can be equivalently rewritten in the form
\begin{displaymath}
W \: = \: \sum_{ij} \phi_i A^{ij}(p) \phi_j
\end{displaymath}
where $A^{ij}$ is a symmetric matrix with entries linear in the $p$'s.
Away from the locus where $A$ drops rank, {\it i.e.}, away from the
hypersurface $\mbox{det }A = 0$, the $\phi_i$ are all massive,
leaving only the $p_i$ massless, which all have charge $-2$.
A GLSM with nonminimal charges describes a gerbe \cite{ps1,ps2,stxglsm},
and physically a string on a gerbe is equivalent via T-duality 
to a string on a disjoint
union of spaces \cite{ps4} (see \cite{ps5} for a short review).

For later use,
let $\Delta$ denote the locus
\begin{displaymath}
\Delta \: = \: \{ \mbox{det }A \: = \: 0 \}
\end{displaymath}
where the mass matrix drops rank.

So far we have found that the Landau-Ginzburg point physics corresponds
to a sigma model on some sort of double cover of ${\bf P}^3$,
away from the hypersurface $\{ \mbox{det }A = 0\} \equiv \Delta$.
The ${\bf Z}_2$ gerbe
on the ${\bf P}^3$ away from $\Delta$ is a banded\footnote{The ${\bf Z}_2$
gerbe on ${\bf P}^3$ is banded, hence the restriction is also banded.
The restriction also should be nontrivial, just as the original
gerbe on ${\bf P}^3$.  Briefly, in light of
\begin{displaymath}
H^2({\bf P}^3,{\bf Z}_2) \: \longrightarrow \:
H^2({\bf P}^3 - \Delta, {\bf Z}_2 ) \: \longrightarrow \:
H^1(\Delta, {\bf Z}_2)
\end{displaymath}
if $\Delta$ is smooth, then it is simply-connected, and so
$H^1(\Delta,{\bf Z}_2) = 0$, which implies that the restriction 
of the gerbe with characteristic class $-1 \mbox{ mod } 2$ is another
nontrivial gerbe on ${\bf P}^3 - \Delta$.
} gerbe and so
\cite{ps4}
gives rise to a disjoint union of two copies of the underlying space,
{\it i.e.} a trivial double cover.
However, we have claimed that we will ultimately get a branched double
cover of ${\bf P}^3$, and 
the branched double cover of ${\bf P}^3$ is a {\it nontrivial}\footnote{
The question of triviality of the
cover is local near $\mbox{det }A=0$, and locally the cover is the
subvariety in ${\bf P}^3 \times {\bf C}$ given by $z^2 = f(x)$,
where $f = \mbox{det }A$, $x$ indicates homogeneous coordinates on ${\bf P}^3$,
and $z$ is a coordinate on ${\bf C}$.  So the cover is trivial if and only
if we can extract locally a square root of $f = \mbox{det }A$.  But if 
$f$ has a square root locally, then it has a square root globally, 
{\it i.e.} $f = g^2$ for some homogeneous polynomial of degree $4$.
So the double cover is trivial if and only if $\mbox{det }A$ is a square,
which usually is not the case.}
double cover of ${\bf P}^3$ away from the branch locus $\Delta$.
The reason for this apparent mismatch is another bit of physics;
to fully understand the Landau-Ginzburg point, we must take into
account a Berry phase, that exchanges the two copies as one circumnavigates
the branch locus, and makes the double cover nontrivial.

\subsection{Berry phase computation}   \label{berryphase}

We shall construct a local model
for the codimension-one 
degenerate locus $\Delta$ so that we can investigate the
fibration structure of the ${\bf Z}_2$ gerbe over the base.

We focus on a smooth point of $\Delta$.
First let us work on the affine patch $p_4 = v \neq 0$.
Choose affine coordinates $z_a \equiv p_a / v, a = 1,2,3$.
The vev of $p_4$ breaks the $U(1)$ spontaneously down to
a ${\bf Z}_2$ subgroup under which the $z$ are even
and the $\phi$ are odd.  The moduli space of the
theory at $r \ll 0$
is parametrized by $z$.  The fields $\phi$ are
massive over a generic point in moduli space.  Redefine
$\phi$ as $y/\sqrt{v}$, so that $v$ drops out of
the superpotential.

Choose
local coordinates so that the
defining equation of $\Delta$ is
$z_3 = 0 + o(z_a^2)$.  Then rescale the $z_a $
by an infinite amount $z_a \to \Lambda^{-2} z_a,
y_i \to \Lambda y_i$, in order to
get rid of the order $z^2_a $ terms
in the defining equation for $\Delta$. 
This flattens out the degenerate locus to a 
hyperplane $z_3 = 0$ in $z_a$ space 

Finally, choose a basis for the $\phi_i$ so that the
matrix $A^{1i}(p) = 0$ and $A^{ij} = m\delta_{ij}$ for
$i,j\geq 2$.  

In the scaling limit where we recover the local model, the
superpotential is
\begin{displaymath}
W = \frac{1}{2} m 
\left( 
z_3 (y^1)^2 + \sum_{i = 2}^8 (y^i)^2
\right)
\end{displaymath}
The $y_i$ for $i > 1$ are massive everywhere in the local model
and decoupled from the $z$ degrees of freedom, so we can
integrate them out trivially.  Likewise the $z_{1,2}$ are
decoupled, massless degrees of freedom parametrizing the
two flat complex dimensions longitudinal to the degenerate
locus.  We shall henceforth ignore them as well.  We are
left with the degrees of freedom $z_3 \equiv z$ and
$y^1 \equiv y$, with superpotential $ \frac{1}{2} m z y^2$,
and a ${\bf Z}_2$ action under which the field $z$ is invariant but
the field $y \mapsto - y$.

Now consider a circle in the $z$ plane surrounding the
degenerate locus $z = 0$.  Treating the
theory as a fibration means doing the path integral in
two steps.  First hold fixed the base coordinate $z$ and
allow $y$ to fluctuate, deriving an
effective theory for $z$.  Then quantize $z$,
with its evolution specified by the
effective Hamiltonian derived in the first step.
This Wilsonian treatment of the path integral
breaks down only in the neighborhood of $z = 0$, where
the $y$ degree of freedom becomes light.  However we can still
ask about the boundary conditions for wavefunctions in a
region defined by removing a disc $D$ containing the
origin of $z$-space.

For values of $z$ in ${\bf C}_z - D$ we know that
because of the noneffective ${\bf Z}_2$ orbifold action,
the fiber theory of $y$ has two degenerate vacua, in one of which
$y$ is untwisted and in the other of which $y$ is twisted.
As was argued in \cite{ps4}, the infrared
limit of the $y$ theory over a given
point in ${\bf C}_z - D$ is equivalent to
a disconnected theory of two discrete points.  So we have
two points fibered over the complement of a
disc in the $z$-plane.  These two points are
defined by the universe operators $U_{\pm} \equiv
\frac{1}{2}(1 \pm \Upsilon)$, where $1$ is the untwisted vacuum and $\Upsilon$
is the twisted vacuum.  

We would like to find out whether the effective theory for $z$ 
defines a trivial or nontrivial fibration over ${\bf C}_z - 
D$.  How can we understand the monodromy of
the two points over the origin?  The two points must
either be exchanged or remain the same as one 
executes a loop in the $z$ plane around the boundary of
the disc $D$.  If the points remain the same, then
$P_{\pm}$ come back to themselves, or equivalently
the twisted vacuum $\Upsilon$
comes back to itself.  If the two fiber points are
interchanged by going around $\partial D$, then that is equivalent to
$P_{\pm}$ being interchanged with $P_{\mp}$, which in
turn is the same as saying that the twisted vacuum $\Upsilon$ comes
back to itself up to a minus sign.

Next, we need to determine whether
$\Upsilon$ comes back to itself with
a $+$ or a $-$ sign when the string is moved 360 degrees
around the boundary of the disc $D$.  Take the
worldsheet to
be compact with radius $r_{ws}$, and the disc $D$ 
in the $z$-plane to have radius $R_D$.  Then
consider contributions to the worldsheet
path integral in which the
string moves around in a circle $|z| = R > R_D$
in a time $T$.   We assume $T \gg r_{ws}$ and also
$T \gg 1/(mR)$.
Without loss of generality we shall also assume $r_{ws} m R \ll 1$,
so that the mass term is important only for the
dynamics of zero modes on the circle and can be
ignored for the nonzero modes.

For any history of $z$ the dynamics of $y$ are exactly 
Gaussian.  That is, the field $y$ and its fermionic
superpartners are controlled by a quadratic but
time-dependent field theory.  Since the field $z$ couples
only in the superpotential, the kinetic term for $y$ 
is $z$-independent, and only its potential is $z$-dependent.

Assume $z$ is independent
of the spatial worldsheet coordinate $\sigma^1$.  Also assume
$z$ stays exactly on the circle $|z| = R$ and
only its phase changes as a function of worldsheet time
$\sigma^0 \equiv t$:
\begin{displaymath}
z \: = \: R \exp(i \omega(t))
\end{displaymath}
Since the mass term for the scalar $y$ in the $Y$
multiplet is $|z|^2 |y|^2$, 
it equals $R^2 |y|^2$ for the particular
$z$-history we consider.  
Thus the phase $\omega$ decouples completely from the 
dynamics of the boson $y$, which is
then just a massive boson which can be integrated out
trivially.

The fermions $\psi_{\pm}^y, \psi_{\pm}^{y\dag}$ do however couple
to the phase of $z$.  Their Yukawa coupling is
\begin{displaymath}
{\cal L}_{Yuk.} \: \equiv \:
m z \psi_-^y \psi_+^y \: - \: m \overline{z} \psi_-^{y\dag} 
\psi_+^{y\dag}
\end{displaymath}
which for our choice of history for $z$ equals
\begin{displaymath}
{\cal L}_{Yuk.} \: = \:
m R 
\left(
\exp(i \omega(t)) \psi_-^y \psi_+^y \: - \:
\exp(- i \omega(t)) \psi_-^{y \dag} \psi_+^{y \dag}
\right)
\end{displaymath}

Since $z_{\sigma^1} = 0$, the $y$-fermion theory is
translationally invariant in the $\sigma^1$ direction.  It is
also Gaussian, so the dynamics factorizes into an infinite product
of finite dimensional Hilbert spaces labelled by spatial
Fourier modes.  That is, if 
\begin{displaymath}
\psi_{\pm(s)}^y \: \equiv \: \frac{1}{2 \pi} 
\int \exp(- i s \sigma^1) \psi_{\pm}^y (\sigma^1)
\end{displaymath}
and similarly for $\psi_{\pm}^{y\dag}$, then the eight operators 
$\psi_{\pm(\pm r)}^y$ and $\psi_{\pm(\pm r)}^{y\dag}$ are
decoupled from all other operators with distinct 
absolute value of $r$.  We are working in the
limit $r_{ws} m R \ll 1$ so for $s\neq 0$ the mass terms
of magnitude $m R$ make a contribution to the frequency
of the oscillators which is negligible compared to the
contribution $s/r_{ws}$ from the spatial gradient.
Therefore the nonzero mode oscillators can never contribute
to the Berry phase.

To completely specify the fermions, we have to specify their boundary
conditions as we traverse the circle.
Without loss of generality, we may assume the fermions are in an NS
sector on the circle.
In the untwisted NS sector, all values of $s$ are half-integral,
so there are no zero modes, and as a result, from the analysis above
there is no contribution to the Berry phase as $\omega$ is
varied from $0$ to $2\pi$.  In the twisted NS sector the
fermions are integrally moded, and so from the analysis above 
there is a contribution to
the Berry phase from the zero mode oscillators
$b_{\pm} \equiv \psi_{\pm(0)}^y$ and $b_{\pm}^{\dag}
\equiv \psi_{\pm (0)}^{y\dag}$.
If we specified that the fermions were in an R sector on the circle,
the analysis would be completely symmetric, just exchanging the
interpretation of twisted and untwisted sectors.

The eigenvalue of the monodromy
on $\Upsilon$ can therefore be obtained by restricting to
zero modes, and so is equivalent to the calculation of the
Berry phase of the system
\begin{displaymath}
H \: \equiv \: m R 
\left(
\exp(i \omega ) b_- b_+ \: - \: \exp(- i \omega) b_-^{\dag}
b_+^{\dag} \right)
\end{displaymath}
as $\omega$ varies from $0$ to $2\pi$.

The result is that the Berry phase on $\Upsilon$ is $-1$.
We can see this as follows.

Represent the fermionic oscillators as gamma matrices:
\begin{eqnarray*}
b_+ & \equiv & \frac{1}{\sqrt{2}} (\Gamma^1 \: + \: i \Gamma^2)
\\
b_- & \equiv & \frac{1}{\sqrt{2}} (\Gamma^3 \: + \: i \Gamma^4)
\end{eqnarray*}
It is clear that the modes then satisfy canonical
anticommutation relations.  

Taking the representation
\begin{eqnarray*}
\Gamma^1 & = & \sigma^1 \otimes \sigma^1
\\
\Gamma^2 & = & \sigma^2 \otimes \sigma^1
\\
\Gamma^3 & = & \sigma^3 \otimes \sigma^1
\\
\Gamma^4 & = & 1 \otimes \sigma^2
\\
\Gamma_{(5)} & = & 1 \otimes \sigma^3
\end{eqnarray*}
we find that
\begin{displaymath}
b_- b_+ \: = \: (\sigma^1 + i \sigma^2) \otimes  \frac{1}{2} (1 + \sigma^3)
\end{displaymath}
so the Hamiltonian is
\begin{displaymath}
H(t)
\: = \: 2 m R \left(
\begin{array}{cc}
0 & \exp(i \omega(t)) \\
\exp(-i \omega(t) ) & 0 \end{array} \right)
\otimes \frac{1}{2}(1 + \sigma^3)
\end{displaymath}

For the rest of the analysis, we will implicitly carry along the
$\otimes (1/2)(1 + \sigma^3)$ as a spectator.
Omitting that factor,
the Hamiltonian is
\begin{displaymath}
H(t) \: = \: 2 m R \left(
\begin{array}{cc}
0 & \exp(i \omega(t)) \\
\exp(-i \omega(t)) & 0
\end{array} \right)
\end{displaymath}

The Berry phase is the eigenvalue of time translation
during the period $[0,T]$ in which $\Delta \omega = 2\pi$, taking
the limit $\dot{\omega} \sim \frac{1}{T} 
\ll 2 m R$.  For this particular system
the limit is unnecessary and the Berry phase is exact
even for $T^{-1}$ comparable with $2 m R$ or large compared to it.
The result is a phase shift given by $\frac{1}{2}(2\pi) = \pi$,
and such a phase shift is equivalent to a sign flip:
$\cos(x + \pi) = - \cos(x)$, $\sin(x + \pi) = - \sin(x)$.

Now let us compute the Berry phase.

Berry's definition of parallel transport is that a state
$\kket\omega$ always be an energy eigenstate as the Hamiltonian
varies through the space of nondegenerate operators, and that
$\d \kket{\psi}$ be orthogonal to $\kket{\psi}$.  Equivalently,
for a set of energy levels $\kket n$, Berry's parallel transport
can be expressed as
\bbb
\d \kket n = \sum_{m\neq n} {{\bbra  m (\delta H) \kket n}
\over{ E_ 0\up n -  E_ 0 \up m}} \kket m
\eee
There are just two energy eigenstates $\kket\pm$ which
always have eigenvalues $E_ 0\up\pm = \pm K$.  The solutions
to these equations are:
\bbb
\kket \pm = \lrdd
\bm { \exp(- \hh i \omega) \cr \cr \pm\exp(+\hh i \omega)} \end{matrix}
 \rrdd
\eee
It is clear that as $\omega \to \omega + 2\pi$, each state gets
a phase of $\pi$, or equivalently a sign of $-1$.
Thus, since we are working in the twisted NS vacuum $\Upsilon$,
we see that $\Upsilon$ gets a Berry phase of $-1$.
The Berry phase arose from fermion zero modes, and there are not any
present in the untwisted NS vacuum $1$, so as noted earlier
the untwisted NS vacuum
does not get any Berry phase.  (R sectors are symmetric.)

We conclude that transporting a pointlike string state\footnote{That
is,
a configuration which is independent of 
the spatial worldsheet coordiante $\sigma^1$.}
around a loop in ${\bf C}_z - D$ which surrounds $D$ once
induces a trivial phase on the untwisted NS vacuum $1$ 
and a phase of $-1$ on the twisted NS vacuum $\Upsilon$.

This will also be true of all bulk NS states, since our
calculation is unaffected by exciting degrees of freedom in 
the $z_{1,2}$ coordinates and their superpartners,
in the 3+1 macroscopic 
Minkowski coordinates $X^{0,1,2,3}, \psi^{0,1,2,3},
\tilde{\psi}^{0,1,2,3}$ of visible spacetime, or even oscillator modes
of the $z_3$ coordinate and its superpartners.  The
zero modes in the angular and
radial $z_3$ directions are what
we have held fixed in order to perform the Berry phase calculation.
Again, R sector states are symmetric.

Since all manipulations above are
entirely local, the calculation holds for any model in which
the degenerate locus is a smooth hypersurface.

Thus, the Landau-Ginzburg point of the GLSM for ${\bf P}^7[2,2,2,2]$
seems to consistently describe a branched double cover of
${\bf P}^3$.  To summarize our progress so far, away from the
branch locus the GLSM at low energies reduces to an abelian gauge
theory with nonminimal charges -- which describes a gerbe,
which physics sees as a multiple cover \cite{ps1,ps2,stxglsm,ps4,ps5}.
The gerbe in question is banded, which would imply a trivial cover, were
it not for Berry phases which wrap the components nontrivially, and so
gives us a nontrivial double cover.

\subsection{Monodromy around the Landau-Ginzburg point}
\label{monbase}

We have discussed how the Landau-Ginzburg point appears to be
describing a nonlinear sigma model on a branched double of ${\bf P}^3$.
In this section we will check that interpretation indirectly
by computing the monodromy about the Landau-Ginzburg point
and showing that it is compatible with a nonlinear sigma model interpretation,
namely, that it is maximally unipotent.

As we will also discuss related monodromy computations
for other models, in this section let us first set up some
generalities.

We consider a Calabi-Yau 3-fold, $X$, with a 1-dimensional K\"a{}hler moduli 
space.  For simplicity, we will take $X$ to be simply-connected. 

Let the generator of $H^2(X)$ be $\xi$.  Then, one topological invariant is 
the positive integer, $p$, such that
\begin{displaymath}
\xi^2 = p \eta
\end{displaymath}
where $\eta$ is the generator of $H^4(X)$. Let $\rho=\xi\eta$ be the 
generator of $H^6(X)$. We obtain another integer, $q$, by writing
\begin{displaymath}
c_2(X)= 2 q \eta
\end{displaymath}
As our basis for $K^0(X)$, we will choose a set of generators, whose ring 
structure mimics that of the even-dimensional cohomology. 
To whit, we will choose
\begin{enumerate}
\item the class $[{\cal O}]$ of the trivial line bundle $\mathcal{O}$
\item $a= [H]\ominus[\mathcal{O}]$, where $H$ is the line bundle with $c_1(H)=\xi$.
\item $[b]$, where $a\otimes a = p b$.
\item $[c]$ where $c = a\otimes b$.

\end{enumerate}
There is a skew-bilinear form on $K^0(X)$, given by
\begin{eqnarray*}
   (v,w) &= & Ind \, \overline{\partial}_{v\otimes\overline{w}}\\
         &= & \int_X ch_3(v\otimes\overline{w}) +\frac{1}{12} (c_1(v)-c_1(w))c_2
(X)
\end{eqnarray*}
Expressed in our basis, this skew-form is represented by the matrix
\begin{displaymath}
\Omega=
\left[ \begin{array}{cccc}
      0&-(p+q)/6&-1&-1\\
(p+q)/6&    0   & 1& 0\\
      1&   -1   & 0& 0\\
      1&    0   & 0& 0
\end{array} \right]
\end{displaymath}
On the mirror, the periods of the holomorphic 3-form obey a Picard-Fuchs 
equation with three regular singular points.

Two of the three monodromies have very simple interpretations in terms 
of operations in K-theory.

The monodromy matrices one extracts from this take a nice form, when thought 
of in terms of natural operations in K-theory. The large-radius monodromy is
\begin{displaymath}
M_\infty: v \mapsto v\otimes H
\end{displaymath}
where $H$ is the hyperplane bundle, corresponding to shifting the $B$ field
as one walks around the large-radius limit in the complexified
K\"ahler moduli space. 
Such a large-radius monodromy is necessarily maximally unipotent, meaning,
\begin{displaymath}
( M_{\infty} \: - \:  1 )^{n+1} \: = \: 0, \: \: \:
( M_{\infty} \: - \: 1 )^p \: \neq \: 0, \: 0 < p \leq n
\end{displaymath}
where $n$ is the dimension of the space (in the present case, $3$),
for the simple reason that in K-theory, we can think of $(M_{\infty} - 1)$ as
tensoring
with $( [H] \ominus [{\cal O}] )$,
and tensoring with $([H] \ominus [{\cal O}])$ is nilpotent -- for example,
$\mbox{ch}( H \ominus {\cal O})^{n+1} = 0$.
Furthermore, if the local coordinates on the moduli space are a cover,
then it might take several turns about the limit point to reproduce all
of $M$, so that in general, the monodromy need merely be maximally
unipotent in the weaker sense that
\begin{displaymath}
( M_{\infty}^N \: - \:  1 )^{n+1} \: = \: 0, \: \: \:
( M_{\infty}^N \: - \: 1 )^p \: \neq \: 0, \: 0 < p \leq n
\end{displaymath}
for some positive integer $N$.

In principle, by checking whether the monodromy about a given point in 
moduli space is maximally unipotent, we can check whether that point
can be consistently described by a nonlinear sigma model on a smooth
Calabi-Yau target.

The monodromy about the (mirror of the) 
conifold is
\begin{displaymath}
M_1: v \mapsto v - (v,{\cal O}){\cal O}
\end{displaymath}
where ${\cal O}$ is the trivial line bundle.
This is the Witten effect, in essence.  In a type II string, an
electrically-charged particle becomes massless at this point, and so
magnetically-charged particles pick up an electric charge proportional
to the effective theta angle, which shifts when one circles the conifold
point.  In these one-K\"ahler-parameter, simply-connected, Calabi-Yau's,
only one species becomes massless:  the wrapped D6-brane.
Of course, the monodromy around $z=0$ is the product of the other two.

In our basis, these monodromies are represented by the matrices
\begin{displaymath}
M_\infty=
\left[ \begin{array}{cccc}
1&0&0&0\\
1&1&0&0\\
0&p&1&0\\
0&0&1&1
\end{array} \right],
\qquad
  M_1=
\left[ \begin{array}{cccc}
1&-(p+q)/6&-1&-1\\
0&1&0&0\\
0&0&1&0\\
0&0&0&1
\end{array} \right]
\end{displaymath}

Now, let us restrict to ${\bf P}^7[2,2,2,2]$.
This is the case $p=16$, $q=32$.

The Picard Fuchs equation for the mirror is
$$
  D \varpi(z) = 0
$$
where $D$ is the differential operator ($\theta-z = z \frac{d}{d z}$):
$$
   D = \theta_z^4 - 16 z (2\theta_z +1)^4
$$
The large-radius point is $z=\infty$. The (mirror of the) conifold is $z=1$,
and our mysterious Landau-Ginzburg point is $z=0$.

In our chosen basis for $K^0(X)={\bf Z}^4$, 
the skew bilinear form, $(\cdot, \cdot)$ is represented
by the matrix
$$
   \Omega \: = \: \left[ \begin{array}{cccc}
               0 & -8 & -1 & -1 \\
               8 &  0 &  1 &  0 \\
               1 & -1 &  0 &  0 \\
               1 &  0 &  0 &  0
            \end{array} \right]
$$
and
$$
 M_\infty \: = \: \left[ \begin{array}{cccc}
               1 &  0 & 0 & 0 \\
               1 &  1 & 0 & 0 \\
               0 & 16 & 1 & 0 \\
               0 &  0 & 1 & 1
            \end{array} \right]
$$
is the large-radius monodromy,
$$
     M_1 \: = \: \left[ \begin{array}{cccc}
               1 & -8 & -1 & -1 \\
               0 &  1 &  0 &  0 \\
               0 &  0 &  1 &  0 \\
               0 &  0 &  0 &  1
            \end{array} \right]
$$
is the conifold monodromy and
$$
   M_0 \: = \: M_\infty M_1 \: = \: \left[ \begin{array}{cccc}
               1 & -8 & -1 & -1 \\
               1 & -7 & -1 & -1 \\
               0 & 16 &  1 &  0 \\
               0 &  0 &  1 &  1
            \end{array} \right]
$$

This last one does not look too illuminating. However:
\begin{enumerate}
\item By an integer change of basis (respecting the quadratic form above),
$M_0$ can be put in the form
$$
   M_0 \: = \: - \left[ \begin{array}{cccc}
               1 & 0 & 0 & 0 \\
               1 & 1 & 0 & 0 \\
               0 & 1 & 1 & 0 \\
               0 & 0 & 1 & 1
            \end{array} \right]
$$
which is minus the large-radius monodromy of the double-cover of ${\bf P}^3$.
\item Using the basis above, we can see the monodromy is maximally unipotent.
$$
  (M_0^2 - 1)^4 = 0, \quad  (M_0^2 - 1)^p \neq 0, 0<p<4
$$
\end{enumerate}

Thus, we see that the monodromy about the Landau-Ginzburg point is
maximally unipotent, and hence compatible with a geometric
interpretation of the Landau-Ginzburg point of this model.

\subsection{A puzzle with a geometric interpretation of
the Landau-\\Ginzburg point}

So far we have described how the Landau-Ginzburg point of the
GLSM for ${\bf P}^7[2,2,2,2]$ describes a branched double cover
of ${\bf P}^3$, branched over a degree eight locus -- the octic
double solid Calabi-Yau threefold.
In particular, we have argued how away from the branch locus, the
Landau-Ginzburg point is a ${\bf Z}_2$ gerbe, which physics sees as a 
double cover, and because of a Berry phase, a nontrivial double cover.
We checked this interpretation by computing the monodromy about the
Landau-Ginzburg point, which we saw is consistent with a 
geometric interpretation.

This seems to be a solid description, but there is a puzzle
in the analysis of the Landau-Ginzburg point that are problematic for
a strict geometric interpretation.
Specifically, the
geometry is singular, but the GLSM (at the Landau-Ginzburg
point) behaves as if it were
on a smooth manifold.

In this section, we will go over this difficulty.
In the next section, we will describe how this problem is resolved,
and simultaneously describe how the relationship between the large-radius
and Landau-Ginzburg points can be understood mathematically.

Again, the problem with an interpretation of the Landau-Ginzburg point as a 
branched double cover is that the CFT does not degenerate
at points where the branched double cover is singular -- the gauged linear
sigma model seems to see some sort of resolution of the branched double
cover.  (We will elaborate on the precise nature of this resolution
later; for the moment, we merely wish to establish the physical
behavior of the CFT.)

Following \cite{WitPhases}, the CFT will be singular at a point in the
target space if there is an extra noncompact branched over that point
in the GLSM.
Now, in the GLSM, the $F$ term conditions in this model can be written
\begin{eqnarray*}
\sum_j A^{ij}(p) x_j & = & 0 \\
\sum_{ij} x_i \frac{ \partial A^{ij} }{ \partial p_k } x_j & = &
0
\end{eqnarray*}
On the branch locus, the first $F$ term condition is trivially satisfied,
but not the second, the second prevents the branch locus from having 
a singularity generically.
Physically, the CFT will only be singular for those vectors $(x_i)$
which are eigenvectors of zero eigenvalue of the matrix
$(A^{ij})$, and also simultaneously eigenvectors of zero eigenvalue of each
matrix $( \partial A^{ij} / \partial p_k )$ for each $p$.
Furthermore, for generic quadrics, there are no such solutions --
as we will see below
the CFT described by the GLSM behaves as if it is describing a smooth space.

Let us compare this to a mathematical analysis.
If the branch locus is described as $\{f(x_1, \cdots, x_n) = 0\}$,
then the
double cover is given by $\{y^2=f(x_1,\cdots,x_n)\}$, and it is straightforward
to check that the double cover $\{y^2=f\}$ will be smooth precisely where
the branch locus $\{f=0 \}$ is smooth.
Thus, geometrically, the branched double cover will be singular only
at places where the surface $\{ \mbox{det }A = 0\}$ is singular,
and for generic quadrics, there {\it will} be singular points on the
branched double cover.  

Thus,
the condition that the hypersurface $\{ \mbox{det }A = 0\}$ be singular,
is {\it different} from the condition for flat directions in the GLSM
that we derived above, and so ultimately as a result,
the GLSM behaves as if it were on a smooth space, whereas the
branched double cover is singular.  

{\bf Global analysis}

Let us now justify the statements made above regarding singularities.

First, let us discuss the singularities (or rather, lack thereof)
in the GLSM.
For the first equation to have a non-trivial                  
solution, $p$ must be in the discriminant of our family of quadrics             
and $x$ must be in the kernel of the matrix $A(p)$. Choose an affine            
chart on ${\bf P}^{3}$ which is centered at $p$. Let $u_{1}$,                
$u_{2}$, $u_{3}$ be the local coordinates in which $p = (0,0,0)$.               
In these terms we have that $A = C_{0} + C_{1}u_{1} + C_{2}u_{2} +              
C_{3}u_{3}$, where $C_{i}$ are constant symmetric $8\times 8$                   
matrices. Note that for a generic choice of quadruple of quadrics          
the family $A$ has a determinant which is not identically zero as a 
function of the $u_{i}$'s.  On the other hand   
the first equation says that there exists a non-zero vector $x$                
such that $C_{0}x =  0$, and the second system of equations says               
that $B_{1}x = B_{2}x = B_{3}x = 0$.  This however implies that $A(u) x   
= 0$ for all $u$, i.e. $\det  A(u) = 0 $ identically in $u$.  This 
gives a contradiction.      

Next, let us turn to the singularities of the branched double cover.
Start with the projectivization             
${\bf P}^{35}$ of the $36$ dimensional vector space of all $8\times          
8$ symmetric matrices.  The space of singular quadrics is a divisor $D 
\subset  {\bf P}^{35}$ - the divisor consisting of all quadrics            
of rank at most $7$. Explicitly                                                 
\[                                                                              
D = \{ [A] \in {\bf P}^{35} | \det A = 0 \}.                                 
\]                                                                              
Our four quadrics span a linear ${\bf P}^{3} \subset                         
{\bf P}^{35}$ and the branch locus is just     
the intersection ${\bf P}^{3} \cap D$.

Now the singularities of the intersection ${\bf P}^{3} \cap D$               
occur at the points where ${\bf P}^{3}$ is not transversal to $D$.           
Note that there are two ways in which this can happen: 
1. when                  
${\bf P}^{3}$ intersects $D$ at a smooth point of $D$ but not                
transversally, and 2. when ${\bf P}^{3}$ passes through a singular           
point of $D$.  
These two types of singularities behave differently:  later when
we discuss homological projective duality, we will see that the          
sheaf of Clifford algebras that we get in the h.p.d. will be locally-free 
at singularities of type 1 and will not be locally-free at                      
singularities of type 2.  So this sheaf will be a sheaf of Azumaya  
algebras on the complement of the points of type 2, {\it i.e.} on this 
complement we will have a gerbe over the double cover.  Across these 
points the sheaf of Clifford algebras gives a noncommutative                   
resolution of the singularities.

Now note that the singular locus of $D$ consists of all quadrics of             
rank at most $6$.  Every quadric of rank exactly $6$ is a cone with 
vertex ${\bf P}^{1}$ over a smooth quadric on ${\bf P}^{5}$.  So            
the dimension of the locus of quadrics of rank $6$ in ${\bf C}^{8}$          
is equal to                                                                     
$\dim Gr(2,8) + \dim S^{2}{\bf C}^{6} = 12 + 21 = 33$ or                     
projectively is equal to $32$.  So $Sing D$ is a non-degenerate  
subvariety of  codimension $3$ in ${\bf P}^{36}$ and so every                
${\bf P}^{3}$ intersects it.                                                 
                                                                                
So the double cover is singular and generically has finitely many               
singularities of type 2.

To better understand this matter, we shall return to the local
model of the branched cover.
We consider a local model of a
geometric singularity of the branched cover,
where the mass matrix for the $y$ degrees
of freedom drops in rank by two.
This occurs when the discriminant
locus $\Delta$ has a surface singularity of
the most generic kind  -- an ordinary 
$A_1$ singularity.  This is described geometrically
by a conifold singularity of the total space of the branched
cover, as we shall see quite directly.  However we will also
establish that the CFT is nonetheless nonsingular; there is
no noncompact branch, even over the point at which
the degenerate locus has an $A_1$ singularity and the
total space has a conifold singularity.

{\bf Setup of the local model}

In this section we shall follow the same notation as in our
analysis of Berry phases in section~\ref{berryphase}.
Two of the six $y$ degrees of freedom, $y_{1,2}$, are
involved in the model in a nontrivial
way.  These are the ones which
are simultaneously massless over the singular
point in the degenerate locus.
Label them $y^{\alpha}$ for $\alpha \in\{ 1,2\}$.
They are coupled to the $z_a$ multiplets through a
$z$-dependent mass matrix which vanishes at
the origin.
The other six $y$'s are massive everywhere.

The superpotential is
\begin{displaymath}
W \: = \: M_{\alpha \beta} (z) y^{\alpha} y^{\beta} \: + \:
\sum_{i = 3}^8
(y^i)^2
\end{displaymath}
We integrate out $y_3,\cdots, y_8$ trivially.

The simplest choice for $M$ which manifests an
$SU(2)$ global
symmetry is
\begin{displaymath}
M_{\alpha \beta}(z) \: \equiv \: \frac{1}{2}
m \epsilon_{\alpha \gamma} \sigma^a_{\gamma \beta} z_a
\end{displaymath}
where $\sigma^a$ are the standard
Pauli matrices.  The $z_a$ transform as a ${\bf 3}$
and the $y_{\alpha}$ transform as a ${\bf 2}$.  Everything else
is a ${\bf 1}$.

In components we have
\begin{displaymath}
W \: = \:
\frac{1}{2} m 
\left( y^1 \: \: y^2 \right)
\left( \begin{array}{cc}
z_1 + i z_2 & -z_3 \\
-z_3 & -z_1 + i z_2 \end{array} \right)
\left( \begin{array}{c}
y^1 \\ y^2 \end{array} \right)
\end{displaymath}
The degenerate locus is given by the equation
\begin{displaymath}
0 \: = \: \det M = -m^2 \left( 
z_1^2 + z_2^2 + z_3^2  \right)
\: = \: - m^2 \sum_a z_a^2
\end{displaymath}
The origin is an
ordinary double point singularity, or $A_1$
surface singularity, of the variety $\Delta
\subset {\bf C}^3$.  That
is, the singularity is locally the quotient
singularity ${\bf C}^2/{\bf Z}_2$.

It is easy to see that a branched cover over ${\bf C}^3$
with branch locus $\{ \sum_a z_a^2 = 0\}$ is
a conifold.  Introduce a fourth variable $u$ and
embed
the cover into ${\bf C}^4$ by the equation
\begin{displaymath}
u \: \equiv \: \pm\sqrt{\sum_a z_a^2}
\end{displaymath}
Defining $u\equiv i z_4$ and squaring
both sides we have the
equation
\begin{displaymath}
z_1^2 \: + \: z_2^2 \: + \: z_3^2 \: + \: z_4^2 \: = \: 0,
\end{displaymath}
which is the defining equation of the undeformed conifold,
in standard form.


{\bf Nonsingularity of the CFT}

Despite the fact that the target space in this local model is geometrically
a singular conifold, the CFT is nonsingular.  \it A fortiori, \rm
this establishes that the theory is inequivalent to the
CFT of the standard conifold, which is singular.
More generally, as we outlined with a global analysis at the beginning
of this section, the GLSM singularities are different from the
geometric singularities.  We will show here that the CFT can be
smooth at a geometric singularity of the branched double cover,
to drive home the distinction.

To see this, it suffices to notice that there is
no noncompact branch at the origin.
The F-term equations for $z_a$ are 
\begin{displaymath}
\epsilon_{\alpha \gamma} \sigma^a_{\gamma \beta}  y^{\alpha} 
y^{\beta} \: = \: 0
\end{displaymath}
One can check directly, component by component, that
this does indeed set both $y^{\alpha} = 0$.  An easier way to see
this is to note that $SU(2)$ is transitive on spinors
of fixed norm and the F-term equations are
$SU(2)$ invariant, so 
either all nonzero values of $y^{\alpha}$ satisfy the F-term equations
or else none of them does.  The former possibility is
obviously not true so the $y^{\alpha}$ must vanish classically,
despite the fact that they both become massless at the origin.

Thus, the geometric singularities of the branched double cover do
not coincide with singularities of the CFT arising at the
Landau-Ginzburg point, which is one problem with the proposal that
the Landau-Ginzburg point flow to a nonlinear sigma model on 
a branched double cover.
We shall resolve this discrepancy in the next section, by arguing that
the technically correct interpretation of the Landau-Ginzburg point is
that it flows to a nonlinear sigma model on a 
`noncommutative resolution' of the branched double cover.
In other words, the branched double cover interpretation will be
correct {\it generically}, but the resulting CFT is not quite globally
the same as a nonlinear sigma model on the branched double cover.

\subsection{Resolution of this puzzle -- new CFT's}

Although the Landau-Ginzburg point seems to be very nearly
equivalent to a nonlinear sigma model on a branched double cover of
${\bf P}^3$, the problem in the last section has made it clear
that such an interpretation can not be completely correct.

In addition, we also have a problem of understanding how to relate
the large-radius and Landau-Ginzburg points geometrically.
Ordinarily, in GLSM's the K\"ahler phases are related by
birational transformations, yet no birational transformation exists
in this case, as pointed out earlier in section~\ref{rev-quadrics}.

We propose that these problems are resolved and understood by virtue
of Kuznetsov's ``homological projective duality'' \cite{kuz1,kuz2,kuz3}.
The homological projective dual of ${\bf P}^7[2,2,2,2]$ is
a ``noncommutative resolution'' of the branched double cover of
${\bf P}^3$ that we have seen.
We shall describe homological projective duality in greater generality
in section~\ref{hpd}, but let us take a moment to review what this
means specifically in this case.

The word `noncommutative' in this context is somewhat misleading.
Kuznetsov's work \cite{kuz1,kuz2,kuz3} and related papers define
spaces by categories of sheaves, and use the
term `noncommutative space' to refer to any space (or other object)
whose sheaf theory yields the defining category.
A noncommutative space could be an ordinary space, an ordinary
space with a flat B field that twists sheaves, or even a Landau-Ginzburg
model.
In particular, a `noncommutative space' need not be associated with
a noncommutative algebra.

In the present case, the noncommutative space that is homological
projective dual to ${\bf P}^7[2,2,2,2]$ is the pair $({\bf P}^3,
{\cal B})$ where ${\cal B} \in Coh({\bf P}^3)$ is the 
sheaf of even parts of Clifford algebras
over ${\bf P}^3$.  (The category that defines this noncommutative space
is the category of coherent sheaves on ${\bf P}^3$ which are also
modules over the sheaf ${\cal B}$.)  This pair $({\bf P}^3,{\cal B})$
defines a pair $(Z, {\cal A})$ where $Z$ is the branched double cover
of ${\bf P}^3$ and ${\cal A}$ is essentially just ${\cal B}$ but
reinterpreted.  In the next section, we shall elaborate on
these structures and also describe how they arise physically
in matrix factorization.

To put this in perspective, this means that the conformal field
theory obtained as the IR limit of the Landau-Ginzburg point of the GLSM,
is not a nonlinear sigma model on a branched double cover,
though it is close.  Rather, it is a new conformal field theory,
that locally on smooth patches behaves like a nonlinear sigma model
on the branched double cover, but in a neighborhood of a singularity,
does something different.  (We will justify this interpretation in
more detail later, and we will leave a more thorough examination of
such new conformal field theories, associated to Kontsevich's notion of
an nc space, to future work.)

This addresses the problem described in the last section.
If we are describing some sort of resolution of 
the branched double cover, rather than the branched double cover itself,
then physical singularities will not coincide with geometric singularities
of the branched double cover.

\subsubsection{Homological projective duality}  \label{hpd}

Homological projective duality is a notion that generalizes all of
the equivalences described here.
It is defined in terms of both the derived categories of the spaces,
and in terms of embeddings into projective spaces:
varieties $X$ and $Y$ equipped with morphisms into the dual
projective spaces $f: X \rightarrow {\bf P}V$, $g: Y \rightarrow
{\bf P}V^*$ ($V$ a vector space) are homologically projective dual
if the derived category of $Y$ can be embedded fully and faithfully
into the derived category of the universal
hyperplane section of $X$ (a subset of $X \times {\bf P}V^*$)
in a certain way.
Homological projective duality was introduced in \cite{kuz1};
it is described for quadrics in \cite{kuz2} and for Grassmannians in
\cite{kuz3}.  

The simplest versions of this correspond to classical duality\footnote{
In other words, for a projective space ${\bf P}V$,
a point in the dual projective space ${\bf P}V^*$ with homogeneous
coordinates $[a_0, \cdots, a_n]$ corresponds to a hyperplane in the
original projective space defined by
\begin{displaymath}
a_0 x_0 \: + \: \cdots \: + \: a_n x_n \: = \: 0
\end{displaymath}
where $[x_0, \cdots, x_n]$ are homogeneous coordinates on
${\bf P}V$.
} between
hyperplanes and points of projective spaces.
For a vector space $V$, the embedding
${\bf P}V[1] \rightarrow {\bf P}V$ defined by the inclusion of 
a hyperplane (degree 1) into its ambient projective space,
is homologically projective
dual to the embedding $\mbox{pt} \rightarrow {\bf P}V^*$
of the dual point into the dual projective space.

More complicated examples can be defined by {\it e.g.} Veronese
embeddings.  Recall
(from {\it e.g.} \cite{harris}[p 23]) that the Veronese map of
degree $d$ is a map ${\bf P}^n \rightarrow {\bf P}^N$ of the form
\begin{displaymath}
[ x_0, \cdots, x_n ] \: \longrightarrow \: [ \cdots, x^I, \cdots ]
\end{displaymath}
where the $x^I$ range over all monomials of degree $d$.
Thus,
\begin{displaymath}
N \: = \: \left( \begin{array}{c} n+d \\ d \end{array} \right) \: - \: 1
\end{displaymath}
In particular, a Veronese map of degree 2 defines an embedding
${\bf P}V \rightarrow {\bf P}\mbox{Sym}^2 V$.
Kuznetsov shows \cite{kuz2} that the double Veronese embedding 
is homologically projective dual
to 
\begin{displaymath}
({\bf P} \mbox{Sym}^2 V^*, {\cal B}_0) \: \longrightarrow \:
{\bf P}\mbox{Sym}^2 V^*,
\end{displaymath}
where
the pair $({\bf P} \mbox{Sym}^2 V^*, {\cal B}_0)$ defines a
`noncommutative' or nc space.  (See appendix~\ref{cy-categories} for an
overview of nc spaces.)  This noncommutative space is defined by sheaves
that are modules over ${\cal B}_0$,
where ${\cal B}_0$ is the sheaf of even parts of Clifford algebras
on ${\bf P}(\mbox{Sym}^2V^*)$:
\begin{displaymath}
{\cal B}_0 \: = \: {\cal O}_{{\bf P}(S^2V^*)} \oplus
\left( \Lambda^2 V \otimes {\cal O}_{{\bf P}(S^2V^*)} \right) \oplus
\left( \Lambda^4 V \otimes {\cal O}_{{\bf P}(S^2V^*)} 
\right) \oplus \cdots
\end{displaymath}

As a practical matter, what arises physically is the induced
action of homological projective duality on linear (hyperplane) sections,
{\it not} precisely bare homological projective duality itself.
Suppose we have dual maps $f: X \rightarrow {\bf P}V$ and
$Y \rightarrow {\bf P}V^*$.  Now, let $L \subset H^0( {\bf P}V, {\cal O}(1))$
be a set of hyperplanes, and define $X_L$ to be the complete intersection of
those hyperplanes with the image of $X$.
Since $L$ is a set of linear forms on ${\bf P}V$, the projectivization
${\bf P}L$ is naturally a linear projective subspace of
${\bf P}V^*$.  Define $Y_L$ to be the intersection of the image
of $Y$ in ${\bf P}V^*$ with ${\bf P}L$.
Kuznetsov proves in \cite{kuz1} that the derived categories of
$X_L$ and $Y_L$ each decompose into several Lefschetz pieces with
one essential last piece in the Lefschetz decomposition.
He also shows that the essential pieces of $X_L$ and $Y_L$ are equivalent.
We will see below that the nc spaces
defined by these essential pieces are exactly the ones related by the
change of phase in the GLSM.

To be specific, let us consider complete intersections of
quadrics.  We have just described the induced action on
{\it hyperplanes}:  to describe the induced action on quadrics,
we must find a way to re-embed so that the quadrics become
hyperplanes\footnote{
A careful reader will note that there is a potential presentation-dependence
problem lurking here.  If homological projective duality is defined
on a choice of linear sections, then different choices, different
ways of rewriting the complete intersection as a complete intersection
of hyperplanes in a projective space, might give rise to different
duals.  This might be partly fixed by a nonobvious uniqueness theorem,
and partly it might correspond to different K\"ahler phases in GLSM's.
We will not attempt to root out this issue here, but instead leave it
for future work.
}, in effect.  Now, a quadric in ${\bf P}V$ is the pullback 
of a linear
polynomial on ${\bf P}\mbox{Sym}^2 V$ under the double Veronese
embedding ${\bf P}W \rightarrow {\bf P}\mbox{Sym}^2 V$.
For example,
if four of the homogeneous coordinates on the target are $x_0 x_1, x_0 x_2,
x_0^2, x_1 x_2$, then the
hyperplane
\begin{displaymath}
(x_0 x_1) \: + \: 3 (x_0 x_2) \: - \: 2 (x_0^2) \: + \: 9 (x_1 x_2)\: = \: 0
\end{displaymath}
in the target ${\bf P}\mbox{Sym}^2V$ is the same as a quadric
hypersurface in ${\bf P}V$.
So, we consider hyperplanes on the image of ${\bf P}V$ in
${\bf P}\mbox{Sym}^2V$, which is equivalent to working with quadrics
on ${\bf P}V$. 

Let us work through a particular example, that of a complete intersection
of quadrics in ${\bf P}^7$.
From the arguments above, let us begin with the double Veronese
embedding ${\bf P}^7 \rightarrow {\bf P}^{35}$, which is dual to
$({\bf P}^{35},{\cal B}_0) \rightarrow {\bf P}^{35}$.
Suppose we have a space $L \subset H^0({\bf P}^{35}, {\cal O}(1))$ of
quadrics of which we wish to take the complete intersection.
Let $X_L$ denote that complete intersection, {\it i.e.},
$X_L = \cap_{q \in L} \{q=0\} \cap {\bf P}^7$.
(For example, if $L$ is four-dimensional, then $X_L$ is the complete
intersection of four quadrics in ${\bf P}^7$, precisely the example
we have been studying in detail so far in this paper.)
Since $L$ is a space of linear forms on ${\bf P}^{35} = {\bf P}\mbox{Sym}^2V$, 
the projectivization
${\bf P}L$ is naturally a linear projective subspace of
${\bf P}^{35} = {\bf P}\mbox{Sym}^2V^*$.
Define the dual linear section $Y_L$ to be the intersection of 
${\bf P}L$ with whatever is embedded in the ${\bf P}^{35} = 
{\bf P}\mbox{Sym}^2V^*$.
In particular, that means $Y_L = ({\bf P}L, {\cal B}_0|_{{\bf P}L})$.
When $L$ is one-dimensional and $X_L$ is just one quadric, then
${\bf P}L$ is a point and $Y_L$ is a point equipped with an nc structure
sheaf which is an even part of a Clifford algebra.
When $L$ is four-dimensional (so that $X_L = {\bf P}^7[2,2,2,2]$),
then $Y_L = {\bf P}^3$, and as we shall see in the next section,
the nc space defined by ${\bf P}^3$ with the restriction of
${\cal B}_0$ is a `noncommutative' resolution of a branched double
cover of ${\bf P}^3$.
Finally, we need to take essential pieces in the derived category,
but since both sides are Calabi-Yau, the essential pieces are the
entire derived category.

A point to which the reader might object
is that the dual spaces obtained are noncommutative spaces,
at least in Kontsevich's sense -- meaning, spaces defined by their
sheaf theory.  In simple cases, the duals will be honest spaces,
but for example when the dual variety is singular\footnote{
It is possible to also small-resolve the singularities of the branched
double cover, but the result is necessarily non-K\"ahler.
For more information on such non-K\"ahler small resolutions see
for example \cite{andrei-ell,andrei-nicadd}.
In any event, for our purposes, this is largely irrelevant,
as we can tell from the sheaf theory -- matrix factorizations in the
UV Landau-Ginzburg model -- that physics really is seeing precisely the
noncommutative resolution, 
a fact that will be described in detail in the next section.
}, then the 
noncommutative space will be a noncommutative resolution of singularities,
matching the underlying variety at smooth points but doing something
different at singular points.
In fact, we shall see in the next section that this behavior matches physics
 -- the Landau-Ginzburg points of these GLSM's have sheaf theory
(defined by matrix factorizations) which precisely matches the sheaf
theory obtained by homological projective duality.
So, physics sees noncommutative spaces; in fact, these GLSM's give a
concrete realization of what it would mean for a string to propagate
on a noncommutative space, in this sense.
Put another way, the CFT's at the Landau-Ginzburg points are, in general,
new types of CFT's -- they look like ordinary nonlinear sigma models
close to smooth points of the branched double cover, but are different
close to singular points.

In the paper \cite{meron}, the relevant homological projective duality
began with the duality between the Pl\"ucker embedding
$G(2,V) \rightarrow {\bf P}(\mbox{Alt}^2 V)$, which was homologically
projective dual to $(\mbox{Pf}, {\cal B}_0) \rightarrow
{\bf P}(\mbox{Alt}^2 V^*)$, where $\mbox{Pf}$ denotes a Pfaffian variety
and ${\cal B}_0$ the sheaf of even parts of Clifford algebras that
defines the structure of a noncommutative space over
$\mbox{Pf}$.  As here, the physically-relevant version of homological
projective duality appearing there was its induced action on
hyperplanes.

We conjecture that K\"ahler phases of GLSMs are related by
homological projective duality.
Unfortunately, it is not possible to check this conjecture at present,
as much more needs to be understood about homological projective duality.
For example, the simplest flop (between small resolutions of the basic
conifold) is known \cite{kuz1}[theorem 8.8]
to work through homological projective duality,
but it is not known whether more general flops are also
related by homological projective duality.

\subsubsection{Noncommutative algebras and matrix factorization}

In this section we shall review some pertinent algebraic
structures arising mathematically in homological projective
duality in this example, and how they can be understood via
matrix factorization.   

Let us begin by reviewing the mathematics \cite{kuz2}
of homological projective
duality in this case.
Consider the complete intersection $X$ of four quadrics in
${\bf P}^{7}$. It is h.p.d. to a non-commutative variety
$(P,{\cal B})$, where $P \cong {\bf P}^{3}$ is the parameter
space for the set of 6-dimensional quadrics that cut out $X \subset
{\bf P}^{7}$, and ${\cal B} \in Coh(P)$ is the sheaf of even
parts of Clifford algebras associated with the universal quadric $\pi
: Q \to P$ over $P$. 
In physics terms, the universal quadric is the GLSM superpotential
$\sum_{ij} \phi_i A^{ij}(p) \phi_j$, and for each point on ${\bf P}^3$
we have a quadric, which defines a metric for which we can associate
a Clifford algebra.
The family of quadrics $\pi : Q \to P$
degenerates along a discriminant surface $\Sigma \subset P$ of degree
$8$.  Equivalently, instead of taking the nc space $(P,\mathcal{B})$,
we can consider the double cover $f : Z \to P$ branched along
$\Sigma$, together with  a sheaf of algebras ${\cal A} \to Z$
for which $f_{*}{\cal A} = {\cal B}$.

Let us take a moment
to understand what happens with the data $(P,{\cal B})$, or
equivalently $(Z,{\cal A})$ in the special situation when the
octic $\Sigma$ becomes singular.
The octic may become singular in two different ways.  First, the
plane
\begin{displaymath}
P \: \subset \: {\bf P}(H^{0}({\bf P}^{7},{\cal O}(2)))
\end{displaymath}
can
become tangent to the discriminant in
${\bf P}(H^{0}({\bf P}^{7},{\cal O}(2)))$.  In this case the
double cover is singular, but the sheaf ${\cal A}$ of algebras on
the double cover $Z$ is a sheaf of Azumaya algebras.  Second, it can
happen that $P$ contains a quadric of corank 2.  In this case the
sheaf of algebras ${\cal A}$ is not locally free at the
corresponding point of the double cover.  So, we get a truly
non-commutative situation.

The structure $(P,{\cal B})$ arises physically via matrix
factorization.  Let us return briefly to the GLSM superpotential,
the `universal quadric' $\sum_{ij} \phi_i A^{ij}(p) \phi_j$.
On the face of it, this describes a hybrid Landau-Ginzburg model,
apparently fibered over $P = {\bf P}^2$.  At each point on
$P$, we have an ordinary Landau-Ginzburg model (in fact, a
${\bf Z}_2$ orbifold) with a quadric
superpotential.  Now, matrix factorization for quadratic superpotentials
was thoroughly studied in \cite{kapli1}.  There, it was discovered 
that the D0-branes in such a Landau-Ginzburg model have a 
Clifford algebra structure.  The D0-branes in a Landau-Ginzburg
model with $n$ fields and a quadratic superpotential give rise
to a Clifford algebra over those $n$ fields, with associated metric
defined by the superpotential.
In the present case, where we have
fibered such Landau-Ginzburg models over $P$, the fibered D0-branes,
or more accurately D3-branes when $P$ is three-dimensional,
will have the structure of a sheaf of Clifford algebras.
(After all\footnote{
The Born-Oppenheimer approximation in this context suggests a theorem
regarding the behavior of matrix factorizations in families,
for which we unfortunately do not yet have a rigorous proof.
}, we can equivalently work in the B model, where the
Born-Oppenheimer approximation for large underlying space becomes exact.)
We can refine this even further.  In our examples, the fibers
are not just Landau-Ginzburg models with quadratic superpotential,
but rather are ${\bf Z}_2$ orbifolds of Landau-Ginzburg models with
quadratic superpotentials, so our fibered D0-branes will have the
structure of a sheaf of even parts of Clifford algebras, as that is
what survives the ${\bf Z}_2$ orbifold.  This 
is precisely the sheaf ${\cal B}$ appearing
mathematically.

Furthermore, as described in \cite{kapli1}[section 7.4], 
all the B-branes in a Landau-Ginzburg model with a quadratic superpotential
are modules over the Clifford algebra, so in particular all the
B-branes in the present case should be modules over the
sheaf of Clifford algebras ${\cal B}$.

Thus, we see that matrix factorization in the hybrid Landau-Ginzburg
model precisely recovers the algebraic structure of homological
projective duality in this example.

\subsection{Summary so far}

We have examined the GLSM for ${\bf P}^7[2,2,2,2]$ in detail.
Before proceeding, let us review what we have found.

First, at the Landau-Ginzburg point of this GLSM, we have found
(via an analysis that requires understanding how stacks and gerbes
enter physics) that at generic points, the theory seems to describe
a branched double cover of ${\bf P}^3$, another Calabi-Yau.
This geometry is realized directly as a branched double cover,
rather than as a complete intersection, which is certainly novel.
Furthermore, the branched double cover and the original
complete intersection ${\bf P}^7[2,2,2,2]$ are not birational to one another.

However, the theory at the Landau-Ginzburg point is not in the same
universality class as a nonlinear sigma model on the branched double cover,
but rather defines a new kind of conformal field theory,
one corresponding to a noncommutative resolution of the space.
This noncommutative resolution is defined mathematically by its sheaf theory,
which we recover physically in matrix factorizations at the Landau-Ginzburg
point of the GLSM.

This structure, this duality between ${\bf P}^7[2,2,2,2]$,
is encoded mathematically in Kuznetsov's homological projective duality
\cite{kuz1,kuz2,kuz3}.
It has been discussed elsewhere \cite{meron} how homological projective
duality explains analogous dualities in nonabelian gauged linear sigma models.
We shall see in the rest of this paper more examples of abelian gauged
linear sigma models exhibiting homological projective duality.

\subsection{Generalizations in other dimensions}

Examples of this form generalize to other dimensions easily.
The complete intersection of $n$ quadrics in ${\bf P}^{2n-1}$ is
related, in the same fashion as above, to a branched double cover
of ${\bf P}^{n-1}$, branched over a determinantal hypersurface
of degree $2n$.  These are Calabi-Yau, for the same reasons as
discussed in \cite{ps4}[section 12.2].
Furthermore, the complete intersections and the branched double
covers are related by homological projective duality\footnote{
To check this \cite{kuzpriv}, note that
the space ${\bf P}^{2n+1}$ in the double Veronese embedding is HP-dual
to the sheaf of even parts of Clifford algebras on the space
${\bf P}^{2n^2+5n+2}$
of all quadrics in ${\bf P}^{2n+1}$.
As a corollary,
the derived category of a complete intersection of $n$ quadrics
in ${\bf P}^{2n+1}$ contains the derived category of (a noncommutative
resolution of) a double covering of ${\bf P}^{n-1}$.
This is discussed in \cite{kuz2}.
}.

In the special case $n=2$, we have elliptic curves at either end
of the GLSM K\"ahler moduli space:
the branched double cover is just the
well-known expression of elliptic curves as branched double
covers of ${\bf P}^1$, branched over a degree four locus.
In fact, the elliptic curve obtained at the LG point is the same
as the elliptic curve at large-radius (though the isomorphism between
them is not canonical).  Technically, this follows from the fact that the
branched double cover of ${\bf P}^1$ is the moduli space of degree 2 line
bundles on ${\bf P}^3[2,2]$, and as such, it is isomorphic after one
chooses a distinguished point on ${\bf P}^3[2,2]$.
In section~\ref{vwex}, we shall see an example in which this particular
example of homological projective duality is essentially fibered
over ${\bf P}^1\times {\bf P}^1$.

In the special case $n=3$, we have K3's at either end of the GLSM K\"ahler
moduli space:
the fact that K3's can be described as double covers branched over
sextic curves, as realized here at the Landau-Ginzburg point,
is described in \cite{gh}[section 4.5], and the relation between
the branched double cover and the complete intersection of
quadrics is discussed in \cite{mukai1}[p. 145].
However, the two K3's obtained at either end of the GLSM K\"ahler
moduli space are not isomorphic:  one has degree 8, the other has
degree 2.

For $n=2, 3$, the branched double cover is smooth, but beginning
in $n=4$ and continuing for higher $n$, the branched double cover is
singular.  For $4 \leq n \leq 7$, the branched double cover has merely
ordinary double points, and for $n > 7$, it has worse singularities.
Already for $n=4$, the branched double cover cannot be globally
resolved into a smooth K\"ahler manifold -- one can perform small
resolutions locally at each ordinary double point, but globally
any set of small resolutions will break the K\"ahler property.
Physically, as we have seen, for $n=4$ physics does not see a 
non-K\"ahler space, but instead sees a `noncommutative resolution,'
an nc space.

\section{Example related to Vafa-Witten discrete torsion}
\label{vwex}

\subsection{Basic analysis}

A more complicated example with analogous properties can be
built as follows.
Consider a complete intersection of two quadrics in the total space
of the projectivization of the vector bundle
\begin{displaymath}
{\cal O}(-1,0)^{\oplus 2} \oplus {\cal O}(0,-1)^{\oplus 2} \: \longrightarrow
\: {\bf P}^1 \times {\bf P}^1
\end{displaymath}
The ambient toric variety can be described by a gauged linear sigma
model with fields $u$, $v$, $s$, $t$, $a$, $b$, $c$, $d$,
and three ${\bf C}^{\times}$ actions, with weights
\begin{center}
\begin{tabular}{c|cccccccc}
& $u$ & $v$ & $s$ & $t$ & $a$ & $b$ & $c$ & $d$ \\ \hline
$\lambda$ & 1 & 1 & 0 & 0 & -1 & -1 & 0 & 0 \\
$\mu$ & 0 & 0 & 1 & 1 & 0 & 0 & -1 & -1 \\
$\nu$ & 0 & 0 & 0 & 0 & 1 & 1 & 1 & 1
\end{tabular}
\end{center}
The complete intersection is formed by adding two more fields
$p_1$, $p_2$, each of weights $(0,0,-2)$ under $(\lambda, \mu, \nu)$.
The D-terms have the form
\begin{eqnarray*}
r_{\lambda} & = & |u|^2 \: + \: |v|^2 \: - \: |a|^2 \: - \: |b|^2 \\
r_{\mu} & = & |s|^2 \: + \: |t|^2 \: - \: |c|^2 \: - \: |d|^2 \\
r_{\nu} & = & |a|^2 \: + \: |b|^2 \: + \: |c|^2 \: + \: |d|^2 \: - \:
2 |p_1|^2 \: - \: 2 |p_2|^2
\end{eqnarray*}

The geometry described above is reproduced when 
$r_{\nu} \gg 0$.  In the phase defined by further demanding
$r_{\lambda} \gg 0$ and $r_{\mu} \gg 0$, $u$ and $v$
form homogeneous coordinates on one of the ${\bf P}^1$'s in the base,
and $s$, $t$ form homogeneous coordinates on the other ${\bf P}^1$.
The fields $a$, $b$, $c$, $d$ form coordinates on the fibers
of the ${\bf P}^3$ bundle formed by projectiving the rank four vector
bundle ${\cal O}(-1,0)^{\oplus 2} \oplus {\cal O}(0,-1)^{\oplus 2}$.

Other phases with $r_{\nu} \gg 0$ give birational models of the same,
related by flops.
For example, consider the case
that $r_{\lambda} \ll 0$ and
$r_{\mu} \gg 0$, then $a$, $b$ form homogeneous coordinates on
one ${\bf P}^1$, 
and
$s$, $t$ form homogeneous coordinates on a second
${\bf P}^1$.  The geometry can still be described
as a ${\bf P}^3$ bundle over ${\bf P}^1 \times {\bf P}^1$, which is true
for all phases with $r_{\nu} \gg 0$.

We discover branched double covers when we consider phases with $r_{\nu} \ll 0$.
Suppose that $r_{\nu} \ll 0$ and $r_{\lambda} \gg 0$, $r_{\mu} \gg 0$.
In this phase, $u$, $v$ form homogeneous coordinates on one ${\bf P}^1$,
$s$, $t$ form homogeneous coordinates on a second ${\bf P}^1$,
and $p_1$, $p_2$ form homogeneous coordinates on a third ${\bf P}^1$.
To fully understand this phase we need to closely examine the superpotential,
which is of the form
\begin{displaymath}
W \: = \: p_1 Q_1 \: + \: p_2 Q_2
\end{displaymath}
where $Q_1$, $Q_2$ are quadratic polynomials in the eight variables
\begin{displaymath}
au, \: av, \: bu, \: bv, \: cs, \: ct, \: ds, \: dt
\end{displaymath}
Let $\gamma_i$ enumerate the four variables $a$, $b$, $c$, $d$,
then the
superpotential can be written
\begin{equation}  \label{modsup1}
W \: = \: \sum_{ij} \gamma_i A^{ij}(p) \gamma_j
\end{equation}
where $A^{ij}$ is a symmetric $4 \times 4$ matrix with entries linear
in the $p$'s and quadratic in combinations of $s$, $t$, $u$, $v$.
This superpotential is manifestly a mass term for the 
$\gamma_i$, so generically the $a$, $b$, $c$, $d$'s will be massive,
except over the locus where the rank of $A^{ij}$ drops.
That locus is defined by $\mbox{det }A = 0$,
and is a degree $(4,4,4)$ hypersurface in $[u,v] \times [s,t]\times
[p_1,p_2]$.
Away from that locus, where the $a$, $b$, $c$, $d$ are massive,
the only fields charged under the third $U(1)$ gauge symmetry are
$p_1$, $p_2$, which both have charge $-2$, so we have a branched double
cover, branched over the locus $\mbox{det }A = 0$,
much as in the previous example.

Other K\"ahler phases with $r_{\nu} \ll 0$ are very similar.
Their descriptions can be obtained by switching the pair
$(u,v)$ with $(a,b)$ and/or switching $(s,t)$ with $(c,d)$.
If we do the former, for example, then we rewrite the superpotential in the
form of equation~(\ref{modsup1}) but with the $\gamma_i$ running over
$u, v, c, d$, and the matrix $A^{ij}$ a symmetric $4 \times 4$ matrix
with entries linear in the $p$'s and quadratic in combinations of
$a, b, s, t$.  This phase then generically is a branched double
cover of ${\bf P}^1 \times {\bf P}^1 \times {\bf P}^1$
(with homogeneous coordinates $[a,b] \times[s,t] \times [p_1,p_2]$
instead of $[u,v]\times [s,t] \times [p_1,p_2]$),
branched over the degree $(4,4,4)$ locus $\{ \mbox{det }A = 0\}$.

This example is believed \cite{kuzpriv129} to be another example,
another physical realization,
of homological projective duality, or rather, for each complete
intersection phase with $r_{\nu} \gg 0$, the corresponding phase
with $r_{\nu} \ll 0$ is believed to be related to the
$r_{\nu} \gg 0$ phase by homological projective duality.

\subsection{Some notes on the geometry}

This particular example is closely related \cite{andreitoappear}
to one discussed in
\cite{vafawitten} in connection with discrete torsion\footnote{At the
time that \cite{vafawitten} was written, discrete torsion was considered
a mysterious degree of freedom, possibly intrinsic to CFT.
Since that time discrete torsion has come to be completely
understood \cite{dt3,dtrev} as
a purely mathematical consequence of defining orbifolds of theories
with $B$ fields, neither mysterious nor intrinsic to CFT.}.
There, recall one started with the quotient
$E\times E \times E / ({\bf Z}_2)^2$, for $E$ an elliptic curve,
and deformed to a space $Y$, describable as a double cover of ${\bf P}^1
\times {\bf P}^1 \times {\bf P}^1$ branched over a singular
degree $(4,4,4)$ hypersurface.

In more detail,
let $X$ be the quotient of a product of 3 elliptic curves by the action of $G =
{\bf Z}_2 \times {\bf Z}_2$, where each non-trivial element of $G$
acts by negation on two of
the elliptic curves, and leaves the third one unchanged.  $X$ can be viewed as
a double cover of ${\bf P}^1 \times {\bf P}^1 \times {\bf P}^1$, 
branched over a
surface $S$ of tri-degree
$(4,4,4)$, highly singular.  

One deforms $X$ by deforming this surface $S$.  
Following \cite{vafawitten}, let us not deform $S$ completely, until it is
smooth, but rather only until one has 64 ordinary double points in $S$.
This does not actually give a complete description of the allowable deformed
branching loci $S$, but we should get that from the next description.  Denote
by $Y$ the typical member of the family of allowable deformations of $X$. It is
a C-Y 3-fold with 64 ODP's, with $\mbox{rk Pic}(Y) = 3$, and it moves in
a 51 dimensional
family. By a result of Mark Gross, $\mbox{Br}(Y) = {\bf Z}_2$.

Next, let us consider the other half of the story. 
Let ${\cal E}$ be the vector 
bundle ${\cal O}(-1,0)^2 
\oplus {\cal O}(0,-1)^2$
on ${\bf P}^1 \times {\bf P}^1$. The projectivization 
${\bf P}{\cal E}$ of ${\cal E}$ is a ${\bf P}^3$ bundle over ${\bf P}^1 
\times  {\bf P}^1$,
and as such it comes with a natural ${\cal O}(1)$.
(Depending on your convention as
to what projectivization means, you may need to take -1 instead of 1 in the
definition of ${\cal E}$.) Let $Z$ be the intersection of two general
sections of ${\cal O}(2)$
in ${\bf P}{\cal E}$.  It is a smooth C-Y 3-fold with
$h^{1,1} =3$, $h^{1,2} = 51$, and the
projection to ${\bf P}^1 \times {\bf P}^1$ exhibits it as a genus one fibration with no section
but with a 4-section.

The analysis of the birational models of $Z$ is 
straightforward, and leads to a picture with ${\bf Z}_2 
\times  {\bf Z}_2$ symmetry,
like a square divided into 4 triangles by the diagonals. There are two sets
of 8 ${\bf P}^1$'s in $Z$ which can be flopped independently,
giving rise to three
more birational models for $Z$ (flop one set, flop the other, or flop both
sets simultaneously).

Next, let us describe how the $Y$'s are related to the $Z$'s. Let $f$ denote the
cohomology class of a fiber of the fibration $Z \rightarrow
{\bf P}^1 \times {\bf P}^1$ (which is an
elliptic curve), and consider the moduli space of stable sheaves on $Z$ of
rank 0, first Chern class 0, second Chern class $f$, and third Chern class 2.
What is meant by this is the moduli problem whose general member is
a torsion sheaf on $Z$, supported on a single fiber, and when this is a smooth
elliptic curve, it should be a line bundle of degree 2 on that elliptic
curve.

Now one can show that this moduli space is precisely one of the $Y$'s, and in
fact there is a very explicit construction of the branch locus $S$ of the
resulting $Y$ in terms of the two quadrics $Q_1$ and $Q_2$ whose intersection
gives a given $Z$.  (Briefly, let the first ${\bf P}^1$ have 
coordinates $(s,t)$,
the second $(u,v)$, in the construction of $Z$. 
Now consider the surface $S$ in
${\bf P}^1(a:b) \times {\bf P}^1(s:t) \times {\bf P}^1(u:v)$
defined by the property that $(a:b, s:t, u:v)$
is in $S$ iff the quadric $aQ_1 + bQ_2$, restricted to the ${\bf P}^3$ 
over $(s:t, u:v)$,
is singular.)

For a given $Z$ and the corresponding $Y$, if we small-resolve the
singularities of $Y$ to form $\overline{Y}$ (which is unfortunately
non-K\"ahler), then there is 
\cite{andrei-ell} an equivalence of derived categories
$D(Z) \cong D(\overline{Y},\alpha)$, where $\alpha$ is the nontrivial
element of $\mbox{Br}(Y)$ and $D(\overline{Y},\alpha)$ denotes the
twisted derived category of $\overline{Y}$.
Presumably, the physically-relevant equivalence is between
$D(Z)$ and the (twisted) derived category of a noncommutative resolution
of $Y$, though such a noncommutative resolution has
not yet been constructed mathematically.

There is some additional mathematical structure which is not realized
in physics.  Each branched double cover of ${\bf P}^1 \times
{\bf P}^1 \times {\bf P}^1$ can be understood as a genus one fibration
in three different ways -- basically, pick any one of the three
${\bf P}^1$'s to be the base of a branched double cover of
${\bf P}^1$ forming an elliptic curve.  (Physically, one of the
${\bf P}^1$'s is distinguished, namely the one defined by the
$p$'s, and moreover, a genus one fibration story does not enter
the physics here at all.)

More information can be found in \cite{andreitoappear}.

\subsection{Relation to ${\bf P}^7[2,2,2,2]$}

Not only is this example analogous to 
${\bf P}^7[2,2,2,2]$, as both involve complete intersections of quadrics,
and at Landau-Ginzburg points describe branched double covers,
but in fact in special cases there is a
quantitative relationship.

Given the complete intersection of two quadrics, we can embed 
in ${\bf P}^1 \times {\bf P}^1 \times {\bf P}^7$.
Specifically, given the eight variables
\begin{displaymath}
au, \: av, \: bu, \: bv, \: cs, \: ct, \: ds, \: dt
\end{displaymath}
which have charge $(0,0,1)$ under $(\lambda, \mu, \nu)$.
In the embedding above, we take these eight variables to be
the homogeneous coordinates on ${\bf P}^7$.
These variables are not independent, but rather obey the two quadric
relations
\begin{eqnarray*}
(au) (bv) & = & (av) (bu) \\
(cs) (dt) & = & (ct) (ds)
\end{eqnarray*}
Thus, what started as a complete intersection of two quadrics in
the total space of 
\begin{displaymath}
{\bf P}\left( {\cal O}(-1,0)^2 \oplus
{\cal O}(0,-1)^2 \right) \: \longrightarrow \: {\bf P}^1
\times {\bf P}^1,
\end{displaymath}
is now a complete intersection of
four quadrics (the two above, plus the two original quadrics)
in ${\bf P}^7$.

This maps to ${\bf P}^7[2,2,2,2]$ implicitly shrinks the 16 rational
curves that are involved in the four flops between different presentations
of the complete intersection of 2 quadrics in the
${\bf P}^3$ bundle on ${\bf P}^1 \times {\bf P}^1$, so as a result,
there are no flops in ${\bf P}^7[2,2,2,2]$.

\subsection{Discrete torsion and deformation theory}

At this point we would like to make an observation regarding discrete
torsion, that is not specific to the particular example we have
discussed so far in this section.

The last remaining unresolved question concerns deformation theory,
namely, how can one explicitly reproduce the results of
Vafa and Witten in \cite{vafawitten}?
In \cite{dt3,dtrev}, the other physically observed characteristics
of discrete torsion, such as its original definition in terms of
phase factors in orbifolds, and its projectivization of group actions
on D-branes, were explicitly derived from the idea that discrete torsion
is defined by group actions on $B$ fields.
The only thing that could not be explicitly derived were the old
results of \cite{vafawitten}, though for those an outline was given:
just as happens for line bundles in orbifolds (and is one way of
understanding the McKay correspondence), perhaps the only way to
consistently deform an orbifold with discrete torsion,
consistent with the orbifold Wilson surfaces, is to add nonzero $H$
flux to exceptional submanifolds, which will play havoc with
supersymmetry, lifting previously flat directions but sometimes also
creating new flat directions.

If, on the other hand, we define spaces through their sheaf theory,
which is the notion at the heart of the nc spaces we see appearing
in {\it e.g.} the CFT at the Landau-Ginzburg point of the GLSM for
${\bf P}^7[2,2,2,2]$, then we have another way of thinking about
this issue.  As has been discussed mathematically in
\cite{agw}, then the infinitesimal moduli should be interpreted as a suitable
Hochschild cohomology, and for the example in 
\cite{vafawitten} the Hochschild cohomology reproduces precisely
the deformation theory seen physically. 

We will not comment further on this matter,
but thought it important enough to warrant attention.

\section{Non-Calabi-Yau examples}   \label{non-cy}

In this section, we will consider six non-Calabi-Yau GLSMs exhibiting
behavior that can be understood in terms of Kuznetsov's homological
projective duality.
Our first three examples, involving GLSMs for
${\bf P}^{2g+1}[2,2]$, ${\bf P}^7[2,2,2]$, and ${\bf P}^5[2,2]$,
have Landau-Ginzburg points that can be interpreted as branched double
covers.  We explain, in the discussion of ${\bf P}^{2g+1}[2,2]$,
subtleties related to the fact that the K\"ahler parameter flows,
and to behavior of Witten indices.

In the second trio of examples, involving ${\bf P}^4[2,2]$,
${\bf P}^6[2,2,2]$, and ${\bf P}^6[2,2,2,2]$, there are additional
complications, stemming from the fact that a branched double cover
of the form one would naively expect can not exist.
We discuss how, instead, to get a geometric interpretation, one
must work in a different cutoff limit where the geometry is interpreted
as a space with hypersurfaces of ${\bf Z}_2$ orbifolds instead of
as a branched double cover.

Curiously, in homological projective duality for complete intersections
of quadrics, there is an even/odd distinction (reflected in the examples
above) which is analogous to the distinction between duals for
$G(2,N)$ for $N$ even and odd in \cite{hori1,kuz3}.

\subsection{Hyperelliptic curves and ${\bf P}^{2g+1}[2,2]$}

A non-Calabi-Yau example of this phenomenon can be obtained as follows.
Consider a gauged linear sigma model describing a complete intersection
of two quadrics in ${\bf P}^{2g+1}$.  (We shall assume $g \geq 1$.) 
The superpotential in this theory can be written 
\begin{displaymath}
W \: = \: p_1 Q_1(\phi) \: + \: p_2 Q_2(\phi) \: = \:
\sum_{ij} \phi_i A^{ij}(p) \phi_j
\end{displaymath}
where the $Q_i$ are the two quadrics, and $A^{ij}(p)$ is a symmetric
$(2g+2)\times (2g+2)$ matrix with entries linear in the $p_a$.
For $r \ll 0$, the $\phi_i$ are mostly massive, away from the
degree $2g+2$ locus $\det A = 0$.  Away from that locus, the only
massless fields are the $p_a$, and as they are nonminimally charged,
they describe a gerbe, which physics sees as a double cover.

So, for $r \gg 0$ (and $g > 1$) we get a positively-curved space,
namely the complete intersection of two quadrics in ${\bf P}^{2g+1}$,
whereas for $r \ll 0$ (and $g > 1$) we get a negatively-curved space,
namely a double cover of ${\bf P}^1$ branched over a degree $2g+2$ locus,
{\it i.e.} a hyperelliptic\footnote{For completeness, let us briefly
repeat the analysis of section~\ref{rev-quadrics} here.
From that section, $K_{S} = \pi^{*}\left(\frac{2k+d}{2}H\right)$
where $k = -2$, the degree of the canonical bundle of ${\bf P}^1$,
and $d = 2g+2$.  Thus, $K_S = (2g-2) \pi^* H$, which is the canonical
bundle of a curve of genus $g$.} curve of genus $g$. 

Before commenting further on the mathematics of this situation,
let us review the physics of this gauged linear sigma model.
For $g=1$, both limits correspond to Calabi-Yau's -- in fact to
elliptic curves.  For $g > 1$, the story is more interesting, as neither
side is Calabi-Yau. 

First note that for $g > 1$, there is an axial anomaly and so the theta
angle is meaningless, so the K\"ahler moduli space is (at best)
one real dimensional.  Furthermore, the singularity near the origin
semiclassically\footnote{
What actually happens after we take quantum corrections into account
is more interesting, and described for cases with vanishing classical
superpotential
in \cite{ilarionronen}.  There, it was argued that the K\"ahler moduli space
does not split apart, but rather extra Coulomb vacua emerge,
and those extra Coulomb vacua fix the problem of mismatched Witten indices
that we discuss momentarily.  It is not completely obvious to the authors
how to extend their results to cases with nonvanishing superpotential.
If we simply ignore the classical superpotential (and there are arguments
that this might be nearly the correct procedure), then the extra
Coulomb vacua are the solutions to the quantum cohomology relation
$\sigma^{2g+1 - 2(2)} = q$.  That would give $2g-3$ extra Coulomb vacua,
which is tantalizingly close to what we need to fix a mismatch
between Witten indices of $0$ and $2-2g$ that we will see shortly.
Since we do not understand how to deal with cases with
nonvanishing classical superpotential, which is the case throughout this
paper, we will not discuss this issue further.
We would like to thank I.~Melnikov for a lengthy discussion of this matter.
}
divides the K\"ahler moduli space into two disconnected halves.

The fact that the K\"ahler moduli space splits rescues us from a problem
with the Witten index.  The Euler characteristic of the genus $g$ curve 
at $r \ll 0$ is $2-2g$.  The Euler characteristic of the $r \gg 0$
complete intersection is different.  
The top Chern class of the tangent bundle of the complete intersection
should be $4$ times the coefficient of $H^{2g-1}$ in
\begin{displaymath}
\frac{ (1+H)^{2g+2} }{ (1+2H)^2 }
\end{displaymath}
(denominator from the two quadric equations, numerator from the Euler
sequence for the tangent bundle to ${\bf P}^{2g+1}$, and the
factor of $4$ from the fact that a general plane
in ${\bf P}^{2g+1}$ intersects the complete intersection in $4$ ($= \mbox{deg }
X$) points).  One can see immediately that the Euler characteristic
of the complete intersection is at least always divisible by four,
whereas the Euler characteristic of the genus $g$ curve obeys no
such constraint.  More, in fact:  one can show the Euler characteristic
of the complete intersection is not only always divisible by four,
but in fact always vanishes.

As a result of the Euler characteristic computations above,
the only time when the Witten indices of the $r \gg 0$ and $r \ll 0$
theories match is when $g=1$, the Calabi-Yau case where the K\"ahler
moduli space is one complex dimensional.
For $g > 1$, the Witten indices do not match -- but since the
K\"ahler moduli space has two distinct components for $g>1$, and there
is no way to smoothly move from one component to the other,
the fact that the Witten indices do not match is not a concern.

As another quick check of the physics, let us discuss how
renormalization group flow behaves in these theories for $g>1$.
The gauged linear sigma model predicts that $r$ will flow towards
$- \infty$, which is consistent with both phases.
For $r \gg 0$, we have a positively-curved space,
so it will try to shrink under RG flow, consistent with the GLSM
computation.  For $r \ll 0$, we have a negatively-curved space,
which will try to expand -- meaning, $|r|$ should increase or,
again, since $r \ll 0$, $r$ will flow towards $- \infty$. 
Thus, we see both phases are consistent with the GLSM prediction
that $r$ will flow in the direction of $- \infty$.

This physics naturally latches onto some corresponding mathematics.
It can be shown \cite{bertrampriv} that 
the moduli of (smooth) complete intersections
of two quadrics in ${\bf P}^{2g+1}$ are naturally isomorphic to the
moduli of hyperelliptic curves of genus $g$.
The isomorphism can be summarized as follows.
Given a smooth quadric $Q$ in ${\bf C}^{2g+2}$, there are two
families of maximal isotropic (Lagrangian) subspaces of $Q$.
Given a pencil of quadrics (which is what one gets with a complete
intersection of two), then the set of maximal isotropic subspaces of
the quadrics in the pencil are a double cover of
${\bf P}^1$ minus the singular quadrics.  The set of singular quadrics
is the intersection of ${\bf P}^1$ with the discriminant locus,
which has degree $2g+2$, and can be described 
in the form $\{ \mbox{det }A = 0\}$ where $A$ is a symmetric
$(2g+2) \times (2g+2)$ matrix linear in the $p$'s, exactly as we
have found physically.  Thus the pencil of quadrics naturally gives
rise to a hyperelliptic curve, and our physical picture of this GLSM has
a natural mathematical understanding.

This example can also be naturally understood in terms of
homological projective duality \cite{kuzpriv129}.
As in the first example we studied in this paper,
the homological projective duality of ${\bf P}^{2g+1}$
and of a sheaf of even parts
of Cliford algebras implies that the derived category of ${\bf P}^1$ branched
in $2g+2$ points embeds fully and faithfully into the derived category
of a complete intersection of $2$ quadrics. 
This is written up in \cite{kuz2}.

\subsection{${\bf P}^7[2,2,2]$}

A complete intersection of three quadrics in ${\bf P}^7$ is an example of
a Fano manifold.

Repeating the same analysis as before, one quickly finds that the
Landau-Ginzburg point of the gauged linear sigma model for
${\bf P}^7[2,2,2]$ is a branched double cover of ${\bf P}^2$, branched
over a degree 8 locus.

Let us check that this is consistent with renormalization group flow
in the GLSM.  As remarked in the last section, if the large-radius
limit is Fano, then the Landau-Ginzburg point in a model with
a one-dimensional K\"ahler moduli space had better be of general type,
to be consistent with the behavior of RG flow in the GLSM.
Applying the results of section~\ref{rev-quadrics}, we see
that the branched double cover obeys 
$K_{S} = \pi^{*}\left(\frac{2k+d}{2}H\right)$
where $H$ is a hyperplane class on the base, $d=8$,
and $k=-3$ for ${\bf P}^2$, so $K_S = \pi^*(H)$, and hence is 
of general type, exactly as needed for consistency.

In fact, just as for ${\bf P}^7[2,2,2,2]$, there is a slight subtlety
in that the branched double cover is singular, at points where the
rank of the quadrics drops by two.  Furthermore, these singularities
do not admit crepant resolutions.  However, the gauged linear sigma model
is nonsingular at those points.

The homological projective dual to ${\bf P}^7[2,2,2]$ is a noncommutative
resolution of the branched double cover above.  Furthermore,
matrix factorizations at the Landau-Ginzburg point trivially reproduce
the sheaf theory that defines the noncommutative resolution.

Thus, as in the Calabi-Yau cases studied, we see that the gauged linear
sigma model is realizing Kuznetsov's homological projective duality.

\subsection{${\bf P}^5[2,2]$}

Repeating the same analysis as before for the gauged linear sigma model
for ${\bf P}^5[2,2]$, one finds a branched double cover of ${\bf P}^1$,
branched over a degree 6 locus.  Unlike the case above, 
this branched double cover is nonsingular generically, so typically
no resolution (noncommutative or otherwise) need appear.
As before, that branched double cover is precisely Kuznetsov's homological
projective dual to ${\bf P}^5[2,2]$.

Let us check that this is consistent with renormalization group flow
in the GLSM.  As remarked earlier, if the large-radius
limit is Fano, as is the
case here, then the Landau-Ginzburg point in a model with
a one-dimensional K\"ahler moduli space had better be of general type,
to be consistent with the behavior of RG flow in the GLSM.
Applying the results of section~\ref{rev-quadrics}, we see
that the branched double cover obeys 
$K_{S} = \pi^{*}\left(\frac{2k+d}{2}H\right)$
where $H$ is a hyperplane class on the base, $d=6$,
and $k=-2$ for ${\bf P}^1$, so $K_S = \pi^*(H)$, and hence is 
of general type, exactly as needed for consistency.

\subsection{Degree 4 del Pezzo (${\bf P}^4[2,2]$)}

A complete intersection of two quadrics in ${\bf P}^4$ is a del Pezzo
of degree four \cite{beauville}[p. 49].
This provides another Fano example to study, though in this case the
physics will be more complicated.

Repeating the previous analysis in this model,
the superpotential in this theory can be written in the form
\begin{displaymath}
W \: = \: p_1 G_1(\phi) \: + \: p_2 G_2(\phi) \: = \:
\sum_{i,j} \phi_i A^{ij}(p) \phi_j
\end{displaymath}
where $G_i$ are the two quadrics, 
and $A^{ij}(p)$ is a symmetric $5 \times 5$ matrix with entries
linear in the $p_i$.
As previously, this superpotential defines a mass term for the $\phi_i$,
except over the locus where the rank of $A$ drops.
So, away from the locus $\det A = 0$, the only massless fields are the
$p_i$, which are of charge $2$ and hence define a ${\bf Z}_2$ gerbe
on ${\bf P}^1$, which physics sees as a double cover.

In other words, the resulting theory in the $r \ll 0$ limit looks like
a double cover of ${\bf P}^1$ branched over the locus $\det A = 0$,
which is degree five.  However, there is no space fitting
that description, 
because to be a branched double cover (here, a covering curve),
the branch locus must be even.
(There is a simple way to see this.  Take a dimension 1 slice through the
base.  If a double cover exists, then we should be able to think of the
intersection of each part of the branch locus with the slice as a 
source of a ${\bf Z}_2$ branch cut.  For all example studied previously,
the branch locus was even degree, and so this is consistent;
but in the odd degree case, then one of the
${\bf Z}_2$ branch cuts has nowhere to end, and so the branched double
cover cannot exist.)

The correct interpretation of this model is slightly more subtle.
It is a ${\bf P}^1$ with a ${\bf Z}_2$ orbifold structure in five points,
which can be thought of locally as the quotient stack of a branched two-fold
covering modulo the natural ${\bf Z}_2$ action on the cover.

To see this, we have to think a bit more carefully about the appearance
of branched double covers in theories with mostly-massive fields.
The ${\bf Z}_2$ gerbe interpretation only holds when minimally-charged
fields have been integrated out:  so, we only have gerbes outside a disk
surrounding the locus where minimally-charged
fields become massless.  The size of that disk is determined by the cutoff
(or other regularization) of the theory.  Outside that disk, we can talk
about gerbes and/or double covers, but inside that disk, there is only a 
single cover.

In the examples studied so far, we have been able to consistently work
in a coupling limit where those disks have shrunk to zero size.
We are running into problems in this example because there cannot be
such a coupling limit here -- there cannot be a global branched double cover
interpretation.

However, there is an alternate limit we can take instead.
Instead of shrinking the disks to zero size, we can expand them to cover
the entire space.  Now, instead of getting a branched double cover,
we get a single cover, with something unusual happening at the
massless locus.

We conjecture that at the massless locus, we have a ${\bf Z}_2$
orbifold structure.
First, branched double covers project onto spaces with
hypersurfaces of ${\bf Z}_2$ orbifolds along the branch locus -- simply
imagine orbifolding the double cover by a ${\bf Z}_2$ that
exchanges the sheets.
(Alternatively, think of the ${\bf Z}_2$ orbifolds as providing
branch cuts.  Although for the branched double cover to make sense requires
that the branch locus have even degree, it is consistent for a space
to have an odd degree hypersurface of ${\bf Z}_2$ orbifolds.)
In effect, we are conjecturing that changing the size of the disk
is moving us from the branched double cover to its underlying space,
which is extremely natural.  Only in the limit that the disks have zero
size or cover the space can we recover an honest geometry; varying the
regularization should interpolate between the two.

Notice that in order for this to be consistent, all of our examples
so far should have an alternate interpretation in which we replace
the branched double cover with a space with ${\bf Z}_2$ orbifolds
-- the description as a space with ${\bf Z}_2$ orbifolds should be
universal, and the description as a branched double cover should only
apply when such a double cover can exist.
This also, we believe to be true.  Mathematically, the branched double
covers described previously have isomorphic sheaf theory to
spaces with even degree hypersurfaces of ${\bf Z}_2$ orbifolds -- the
two descriptions are more or less isomorphic from the perspective of
homological projective duality.  We should be able to flip between them
simply by varying the regularization of the theory.
In the Calabi-Yau case, where one gets a nontrivial CFT, one should be
able to vary a cutoff without fundamentally changing the theory,
and so the result should follow from that.  In the non-Calabi-Yau
cases discussed previously, the matter is more delicate.

We still have not directly addressed the
reason why there should be a hypersurface of ${\bf Z}_2$ orbifolds.
In principle, one ought to be able to deduce this from the Berry phase,
but we have not yet worked out a consistent generalization of the
computation described earlier.  Part of the computation lies in the
fact that we are no longer describing nontrivial CFT's in the IR limit,
so the state-operator correspondence is complicated.
Another part of the complication lies in the Berry phase itself,
which should now get a contribution from the curvature of the space -- so
for example, on a ${\bf P}^1$ with a single ${\bf Z}_2$ orbifold singularity
at a point, there should be a nontrivial phase for a loop that closely wraps
the point, but as one expands the loop around the space, there should be
an additional curvature-dependent contribution so that ultimately,
for a loop closely wrapping another ordinary point far from the
${\bf Z}_2$, the total phase vanishes.

Instead of appealing to the Berry phase, we shall appeal to
homological projective duality.  The matrix factorization argument
we discussed in the context of ${\bf P}^7[2,2,2,2]$ naively applies
here (though a B twist is no longer well-defined, we can still
speak about certain D-branes in the untwisted theory), and so can
deduce what sheaf theory the dual theory possesses,
from which we get that the defining sheaf theory is the same as that
for homological projective duality.
Put another way, homological projective duality tautologically matches
the open string sector, as in previous examples, from which we conclude
that it had better also match the closed string sector.

In the present case, the homological projective dual\footnote{
Technically, the HP-dual of ${\bf P}^4$ in the double Veronese
embedding is ${\bf P}^{14}$ with the sheaf of even parts of Clifford
algebras. Therefore, the derived category of a complete intersection
of 2 quadrics in ${\bf P}^4$ contains the derived category of ${\bf P}^1$ with
a ${\bf Z}_2$-orbifold structure in 5 points.
} \cite{kuzpriv129} of 
${\bf P}^4[2,2]$ is a ${\bf P}^1$ with 5 ${\bf Z}_2$ orbifold points,
so we conclude that the Landau-Ginzburg point can be described in this
fashion.
We can no longer claim to give a completely independent check of
homological projective duality in this example, as we cannot give a 
completely physically rigorous derivation of the origin of the
${\bf Z}_2$ orbifold points, but at least we can see how this
is consistent.

We have discussed previously how the GLSM K\"ahler moduli space
semiclassically falls apart into disjoint pieces in 
non-Calabi-Yau cases, and how in one-dimensional examples, if one
limit is Fano, the other must be of general type, {\it i.e.} have
opposite curvature in order for the running of the Fayet-Iliopoulos
parameter in the GLSM to be consistent.

The same analysis applies here, and the ${\bf P}^1$ with 5 ${\bf Z}_2$
orbifolds is of general type, so the Fayet-Iliopoulos RG analysis
discussed previously is consistent.  Intuitively,
the ${\bf Z}_2$ orbifolds have enough negative curvature to counterbalance
the positive curvature of the rest of the projective space.

Let us examine this issue in more detail, for later use.
(The analysis will be closely analogous to that for double covers
described in section~\ref{rev-quadrics}.)
Given an $n$-dimensional projective space and 
$D \subset {\bf P}^n$ a smooth hypersurface of degree $d$,
then consider a stack describing ${\bf P}^n$ with a ${\bf Z}_2$
orbifold along $D$, which shall be denoted
$Z \equiv {\bf P}^n \langle D/2 \rangle$, together with its
natural map $\pi: Z \rightarrow {\bf P}^n$.
On this stack there is a unique smooth divisor
$\widetilde{D} \subset Z$ such that
$\pi^* {\cal O}_{{\bf P}^n}(-2D) = 
{\cal O}_Z(-\widetilde{D})$.
Now, a straightforward computation shows that\footnote{
Note that $d$ need not be even: 
although ${\cal O}(-d/2+n+1)$ does not make sense on the
projective space, the line bundle ${\cal O}(-d + 2n + 2)$ pulls back
to a line bundle on $Z$ which has a canonical square root.
We can then think of this square root as the pullback of the fictional
line bundle ${\cal O}(-d/2 + n + 1)$, and this agrees with the pullback
map on ${\bf Q}$-divisors.  The fictional bundle
${\cal O}(-d/2 + n + 1)$ on the projective space also has an interpretation
as a parabolic line bundle whose parabolic structure is specified along
the divisor where the orbifold structure lies; see
for example \cite{simpson} and references therein
for more information on parabolic vector bundles.
}
$K_Z = \pi^* K_{{\bf P}^n}(\widetilde{D})$,
{\it i.e.} $K_Z = \pi^* {\cal O}(d/2 - n - 1)$ or
$c_1(TZ) = \pi^* {\cal O}(-d/2 + n + 1)$.

As a consistency check, recall that elliptic curves can be
described as branched double covers of ${\bf P}^1$, which project
to ${\bf P}^1$ with 4 ${\bf Z}_2$ orbifold points: 
$d=4$, $n=1$, and so in this case, $K_Z = {\cal O}$,
as one would expect for a Calabi-Yau.
Similarly, some K3's can be expressed as branched double covers
of ${\bf P}^2$, branched along sextics, which project to
${\bf P}^2$ with a ${\bf Z}_2$ orbifold structure along a hypersurface
of degree 6:  $d=6$, $n=2$, and so again we find
$K_Z = {\cal O}$, as one would expect for a Calabi-Yau.
If $d/2 < n+1$, then the projective space with a hypersurface of
${\bf Z}_2$ orbifolds is Fano; if $d/2 > n+1$, then it is of
general type.

In the present case, a ${\bf P}^1$ with 5 ${\bf Z}_2$ orbifold points,
$d=5$ and $n=1$, so $d/2 > n+1$, and so this stack is of general type,
as claimed earlier.

This phenomenon has appeared previously in the physics literature
in discussions of nonsupersymmetric orbifolds, see
for example \cite{evaallan,greg1,greg2}.
There, it was discovered in (nonsupersymmetric) complex codimension one
${\bf Z}_2$ orbifolds that relevant operators in twisted sectors generically
get vevs under RG flow -- if we think of those ${\bf Z}_2$ orbifolds as
having negative curvature, then under RG flow they should `expand'
(meaning, twisted sector operators get vevs).

\subsection{${\bf P}^6[2,2,2]$}

Following the same analysis as in the last section,
the Landau-Ginzburg point in this model should be 
a ${\bf P}^2$ with a degree 7 hypersurface of ${\bf Z}_2$ orbifolds.
As before, since the hypersurface is of odd degree,
there cannot be a branched double cover, but there can be a space
with an odd degree hypersurface of ${\bf Z}_2$ orbifolds, so we
propose that that is the geometry that emerges in a relevant
regularization limit, replacing the massless locus with 
${\bf Z}_2$ orbifolds.

As before, the homological projective dual to 
${\bf P}^6[2,2,2]$ is exactly this, a 
${\bf P}^2$ with a degree 7 hypersurface of ${\bf Z}_2$ orbifolds
\cite{kuzpriv310}.

Let us also check that this relation is consistent with renormalization
group flow in the GLSM.
At large radius, ${\bf P}^6[2,2,2]$ is Fano, so the opposite end of the
one-dimensional GLSM K\"ahler moduli space had better be of general type,
in order to be consistent with the one-loop renormalization of the
Fayet-Iliopoulos parameter in the GLSM.
Indeed, using the results of the previous section,
for a ${\bf P}^2$ with a degree 7 hypersurface of
${\bf Z}_2$ orbifolds, we have $d=7$, $n=2$, and so
$d/2 > n+1$, hence the ${\bf P}^2$ with hypersurface of
${\bf Z}_2$ orbifolds is of general type, exactly as needed to be
consistent with physics.

\subsection{${\bf P}^6[2,2,2,2]$}

Following the same analysis as previously, the Landau-Ginzburg
point in this model should be a ${\bf P}^3$ with a degree 7
hypersurface of ${\bf Z}_2$ orbifolds.
As before, since the hypersurface is of odd degree, there cannot be
a branched double cover, but there can exist globally a space with
a hypersurface of ${\bf Z}_2$ orbifolds.
Also as before, this is also precisely the result of applying
homological projective duality \cite{kuzpriv310}.
In addition, the surface has several ordinary double points,
which are resolved via a noncommutative resolution, of the form
discussed previously.

Let us also check that this relation is consistent with renormalization
group flow in the GLSM.
At large radius, ${\bf P}^6[2,2,2,2]$ is of general type,
so the opposite end of the one-dimensional GLSM K\"ahler moduli space
had better be Fano,
in order to be consistent with the one-loop renormalization of the
Fayet-Iliopoulos parameter in the GLSM.
Indeed, using our earlier results,
for a ${\bf P}^3$ with a degree 7 hypersurface of ${\bf Z}_2$ orbifolds,
we have $d=7$, $n=3$, and so $d/2 < n+1$, hence the
${\bf P}^3$ with hypersurface of ${\bf Z}_2$ orbifolds is Fano,
exactly as needed to be consistent with physics.

\section{More general complete intersections}    \label{genl-ci}

So far in this paper we have discussed exclusively gauged linear
sigma models for complete intersections of quadrics and the
role that homological projective duality plays in such models.
Given that plus the previous work
\cite{meron}, the reader might incorrectly conclude
that homological projective duality can only be applied to GLSM's with
relatively exotic physics.

To counter such a perception, in this section we will describe
a few examples of GLSM's for complete intersections of higher-degree
hypersurfaces, and how their K\"ahler phase structure can also
be (correctly) understood with homological projective duality.

To a limited extent, this was also described in the context of
nonabelian GLSM's in \cite{meron}, where homological projective duality
was used to predict that the Landau-Ginzburg point of
$G(2,6)[1^6]$ should be a Landau-Ginzburg model corresponding to a
K3 surface.  However, this was a prediction, not an observation,
and moreover, Landau-Ginzburg models for K3's have the same CFT
as nonlinear sigma models on K3's, so, the examples in this section
should clarify that homological projective duality really can 
generate true Landau-Ginzburg theories.

\subsection{${\bf P}^4[3]$}

In this section we will study another example of a GLSM for a
higher-degree hypersurface.

The Landau-Ginzburg point of this gauged linear sigma model
is well-known to be a theory defined by Landau-Ginzburg model
on $[{\bf C}^5/{\bf Z}_3]$ with a cubic superpotential.

According to \cite{kuz3} we can recover 
the same result from homological projective duality.
We begin with the embedding
$G(2,6) \rightarrow {\bf P}^{14} = {\bf P} \mbox{Alt}^2 {\bf C}^6$
and its homological projective dual
$\mbox{Pf} \rightarrow {\bf P} \mbox{Alt}^2 {\bf C}^{6*}$ where
$\mbox{Pf}$ is the Pfaffian cubic hypersurface.  Now, consider the
action induced on linear sections.
Let $L$ be a space of linear forms of dimension $5$, so that
${\bf P}L \subset {\bf P} \mbox{Alt}^2 {\bf C}^{6*}$ has dimension 4.
Intersecting both sides, we find that
$G(2,6)[1^5]$ is dual to a cubic threefold, {\it i.e.} ${\bf P}^4[3]$.

Finally, we must consider the essential pieces in each derived
category.  The nc spaces defined by those essential pieces are what
should be appearing at the $r \rightarrow \pm \infty$ limits of
the GLSM K\"ahler moduli space.  
Kuznetsov argues that there is a semi-orthogonal decomposition
\begin{displaymath}
D^b( G(2,6)[1^5] ) \: = \: \langle {\cal O}, S^*,
\Phi \rangle
\end{displaymath}
(where $S$ is the universal subbundle on the Grassmannian,
restricted to the complete intersection)
and
\begin{displaymath}
D^b( {\bf P}^4[3] ) \: = \: \langle {\cal O}, {\cal O}(1), \Psi \rangle
\end{displaymath}
where $\Phi$ and $\Psi$ are the `essential pieces' of the respective
categories.  Kuznetsov also checks that $\Phi = \Psi$.

Homological projective duality does not itself tell
us how to interpret the nc spaces defined by the essential pieces
$\Phi$, $\Psi$ above.
Instead, to interpret the essential piece of the derived
category of ${\bf P}^4[3]$, we turn to work of Orlov 
\cite{orlov1,orlov2,orlov3},
which related matrix factorizations in Landau-Ginzburg models on
$[ {\bf C}^5/{\bf Z}_3 ]$ to sheaves on ${\bf P}^4[3]$.
Specifically, he argued that the category of matrix factorizations
on the Landau-Ginzburg model, which we shall denote
$D^b([ {\bf C}^5/{\bf Z}_3], W)$ embeds into $D^b( {\bf P}^4[3] )$,
as the right orthogonal to the objects ${\cal O}$, ${\cal O}(1)$.
In other words, 
\begin{displaymath}
D^b( [{\bf C}^5/{\bf Z}_3], W) \: = \: \Psi \: = \: \Phi
\end{displaymath}
Thus, the nc space that is homologically projective dual to
${\bf P}^4[3]$, is the Landau-Ginzburg model on 
$[ {\bf C}^5/{\bf Z}_3 ]$ with cubic superpotential,
exactly as happens in the GLSM.

Thus, the nc space which is homologically projective dual
to ${\bf P}^4[3]$, is exactly the Landau-Ginzburg model 
appearing at the Landau-Ginzburg point in the 
GLSM K\"ahler moduli space, giving us another check that
homological projective duality seems to be correctly encoding
the phases of GLSM's, and not just in physically exotic cases.

In this particular example, homological projective duality acts
somewhat trivially:  it automatically preserves nc spaces
(corresponding to the physics that the B model is independent of
K\"ahler moduli), but typically exchanges different presentations of 
a fixed nc space.  In the present case, homological projective duality
did not itself produce a different presentation; instead, we had to
invoke results of Orlov to interpret the nc space.
What is important about this example is the following:
\begin{enumerate}
\item First, although Orlov's work was used more crucially than
homological projective duality, homological projective duality did
not produce a contradiction.  This is more or less guaranteed by the
structure of homological projective duality -- nc spaces are preserved,
and as nc spaces are defined by the open string sector of the B model,
which is K\"ahler-moduli-independent, this is a must.
\item Second, this means that these ideas encompass not only
geometries appearing at limit points in
GLSM K\"ahler moduli spaces, but also theories that one usually thinks of
as being non-geometric.
In the past, many researchers have spoken of the {\it geometric} phases
being related by birational transformation; now we see that the replacement
for `birational' encompasses more than just geometric limit points.
\end{enumerate}

\subsection{${\bf P}^5[3,3]$}

In this section we will study another example of a GLSM for a
higher-degree hypersurface.
Again, we will see that homological projective duality encompasses
not only limits in the GLSM K\"ahler moduli space with geometric
interpretations, but non-geometric points as well.

\subsubsection{Basic analysis}

The Landau-Ginzburg point of this model is what is sometimes called
a `hybrid Landau-Ginzburg model' or `fibered Landau-Ginzburg model.'
It describes what might be very roughly called a family of Landau-Ginzburg
models (over the vector space ${\bf C}^6$) fibered in some fashion over
${\bf P}^1$.

Let us first make this description more precise.
A Landau-Ginzburg model is defined by a holomorphic function 
(the superpotential) over some noncompact space or stack.
In the present case, the noncompact space is defined by a GIT
quotient of six fields $\phi_i$ and two fields $p_a$ by a ${\bf C}^{\times}$
with weights 
\begin{center}
\begin{tabular}{cc}
$\phi_i$ & $p_a$ \\ \hline
-1 & 3 \\
\end{tabular}
\end{center}
If the weights were reversed, so that the $\phi_i$ had weight $-3$ and the
$p_a$ had weight $1$, then this space would be the total space of a rank
six vector bundle over ${\bf P}^1$, namely ${\cal O}(-3)^{\oplus 6}
\rightarrow {\bf P}^1$.  However, something more unusual is going on instead.

A quotient ${\bf C}^N//{\bf C}^{\times}$ where all the fields have
weight $k$ rather than weight 1 is seen physically not precisely as a copy
of ${\bf P}^{N-1}$, but rather as a ${\bf Z}_k$ gerbe over ${\bf P}^1$
\cite{ps1,ps2,stxglsm,ps4,ps5}.
The quotient above is therefore not a vector bundle over ${\bf P}^1$,
but rather is a vector bundle over a ${\bf Z}_3$ gerbe on ${\bf P}^1$.
If the $\phi_i$ all had the same weight as the homogeneous coordinates
on the underlying stack, then we would naturally think of each $\phi$ as
a coordinate on the total space of the
line bundle ${\cal O}(-1)$, but instead the $\phi_i$'s all have charge
equal to a third of that of the underlying stack.
Such a line bundle is conventionally denoted ``${\cal O}(-1/3)$,''
a notation that only makes sense for gerbes and stacks, not spaces.
(Bundles on gerbes will be discussed in greater detail in the
upcoming work \cite{metonyinprogress}.)

Intuitively, this description of the bundle encodes the fact
that we are fibering a ${\bf Z}_3$ orbifold over a base space
 -- the ${\bf Z}_3$ acts trivially on the base ${\bf P}^1$
(and so, we have a gerbe structure), but nontrivially on the fibers.
To give a little more perspective on this language,
the total space of the line bundle ${\cal O}(1/k)$ over
the ${\bf Z}_k$ gerbe on a point (denoted $B{\bf Z}_k$)
is the same thing as the orbifold $[ {\bf C} / {\bf Z}_k ]$.

In any event, we see that
this Landau-Ginzburg model is defined by a superpotential over
the noncompact stack given by the total space of  
the bundle ${\cal O}(-1/3)^{\oplus 6} \rightarrow
G_3 {\bf P}^1$, where $G_3 {\bf P}^1$ denotes the gerbe on ${\bf P}^1$.

The fibers of this hybrid Landau-Ginzburg model look like
Landau-Ginzburg models with a cubic superpotential in six variables.
At least at the Fermat point, such a superpotential splits into two
copies of a cubic superpotential in three variables, or put another way,
$(T^2 \times T^2)/{\bf Z}_3$.  
This suggests that we should think of the theory at the Landau-Ginzburg
point as some sort of K3 fibered over a ${\bf P}^1$ base,
which we will indirectly check in the next section by studying monodromy
computations.
In the section after that, we shall see that homological projective
duality will make the same prediction.

\subsubsection{Monodromy computation}

In this section we shall check the intuition developed in the last
section, that the Landau-Ginzburg point of the GLSM for
${\bf P}^5[3,3]$ should have the structure of a fibration
over ${\bf P}^1$, via a monodromy computation.

We shall follow the conventions of section~\ref{monbase}.
This is the case: $p=9$, $q=27$. The total Chern class is
\begin{displaymath}
c(X)= 1+54\eta-144\rho
\end{displaymath}
which yields $h^{2,1}=71$.

The intersection form,
\begin{equation}
\omega=
\left[ \begin{array}{cccc}
                0&-6&-1&-1\\
                6& 0& 1& 0\\
                1&-1& 0& 0\\
                1& 0& 0& 0
 \end{array} \right]
\label{533omega}\end{equation}
The large-radius monodromy is
\begin{equation}
M_{\infty}: \: v\mapsto V\otimes H =
\left[ \begin{array}{cccc}
                1& 0& 0& 0\\
                1& 1& 0& 0\\
                0& 9& 1& 0\\
                0& 0& 1& 1
\end{array} \right]
\label{533Mr}\end{equation}
The conifold monodromy is
\begin{equation}
M_1: \: v\mapsto V- (v,{\cal O})v =
\left[ \begin{array}{cccc}
                1&-6&-1&-1\\
                0& 1& 0& 0\\
                0& 0& 1& 0\\
                0& 0& 0& 1
\end{array} \right]
\label{533Mc}\end{equation}
The ``{}hybrid''{} monodromy is
\begin{equation}
M_0^{-1} = (M_{\infty} M_1)^{-1} =
\left[ \begin{array}{cccc}
                -5& 6& 0& 1\\
                -1& 1& 0& 0\\
                 9&-9& 1& 0\\
                -9& 9&-1& 1
\end{array} \right]
\label{533Mh}\end{equation}
This does not look very illuminating. But note that the
standard 2-dimensional irreducible representation of ${\bf Z}_3$
\emph{over the integers} (this representation is, of course,
reducible over ${\bf C}$) is generated by
\begin{displaymath}
\gamma = \left[ \begin{array}{cc}-1&-1\\ 1&0\end{array} \right]
\end{displaymath}
In fact, $M_0^{-1} = A M A^{-1}$, where
\begin{equation}
M \: = \:
\left[ \begin{array}{cccc}
                 -1&-1& 0& 0\\
                 1& 0& 0& 0\\
                 0& 1&-1&-1\\
                 0& 0& 1& 0
\end{array} \right]
\label{533M}\end{equation}
where $A\in SL(4,{\bf Z})$. But for the extra ``{}1''{} in the lower
left-hand $2\times2$ block, this would simply be two copies of the
two-dimensional irrep of ${\bf Z}_3$, and we would have $M^3=1$.
Instead, it is unipotent,
\begin{displaymath}
(M^3-1)^2 = 0
\end{displaymath}

If the theory at the Landau-Ginzburg point were an ordinary Landau-Ginzburg
orbifold (here presumably a ${\bf Z}_3$ orbifold due to the
fact that we are intersecting cubics), then because of the
${\bf Z}_3$ quantum symmetry, we would expect the monodromy to
obey $M^3 = 1$.
Instead, we have something different.  The typical interpretation
of a monodromy of the form above, $(M^3 - 1)^2 = 0$, is that we
have a fibered Landau-Ginzburg orbifold,
fibered over a ${\bf P}^1$ base.  Heuristically, 
going around three times is equivalent to shifting the $B$ field on the
${\bf P}^1$.

Thus, we have confirmation from the monodromy computation that physically,
the Landau-Ginzburg point of the GLSM for ${\bf P}^5[3,3]$ should
look like fibered Landau-Ginzburg models, fibered over a
${\bf P}^1$ base, which is what we argued in the last section..
In the next section, we will see that homological projective
duality predicts mathematically that the Landau-Ginzburg point
should be interpreted as a noncommutative K3 fibration, fibered over
a ${\bf P}^1$ base.

\subsubsection{Homological projective duality and fibered noncommutative K3s}
\label{hpdnck3}

Mathematically, the homological projective dual of the complete
intersection ${\bf P}^5[3,3]$ is \cite{kuzpriv129} 
a family of noncommutative K3s, fibered
over ${\bf P}^1$, the same structure that we have
seen physically.  Let us review the mathematical argument
below.

(1)  First, one can use the Grassmannian-Pfaffian homological projective duality
to argue that if $D \subset {\bf P}^5$ is a four-dimensional smooth Pfaffian
cubic hypersurface, then the dual linear section of the Grassmannian is
a K3 surface $S$ of degree $14$.  In this case,
$D^b(S)$ embeds as a full subcategory in $D^b(D)$ and $D^b(D)$ has an induced
semi-orthogonal decomposition in which $D^b(S)$ is left-orthogonal
to $\langle {\cal O}(-3), {\cal O}(-2), {\cal O}(-1) \rangle$.

(2) Taking the observation above as a model, take any smooth four-dimensional
cubic hypersurface $D \subset {\bf P}^5$ (not necessarily Pfaffian) and look at the
left orthogonal ${\cal A}_D \equiv {}^{\perp} \langle
{\cal O}(-3), {\cal O}(-2), {\cal O}(-1) \rangle$.

Now, we can interpret the category ${\cal A}_D$ to be a category of B-branes
on some $c=6$ SCFT, which from a mathematical point of view defines 
(the open string sector of) a nonlinear
sigma model on
a `noncommutative\footnote{Recall from our earlier discussion that
the usage of `noncommutative' in this context
does not refer to any sort of noncommutative algebra or ring.
Rather, `noncommutative space' is the generic term for the base space
when interpreting a category as a category of A or B model branes.
The base space might be a Calabi-Yau, a Calabi-Yau with a flat $B$ field,
a nongeometric Landau-Ginzburg model, or something else.  }
K3.'
The first justification is
(1) where we see that ${\cal A}_D$ specializes to a Pfaffian cubic.  The
second is that ${\cal A}_D$ is a smooth proper Calabi-Yau category\footnote{
See appendix~\ref{cy-categories} for a review of Calabi-Yau categories.} 
of dimension two.
The third is that the Hochschild homology of ${\cal A}_D$ has the same exact
size as the de Rham cohomology of a K3.

(3) Now let us consider families.  If you have $X$ which is the
complete intersection of two cubics in ${\bf P}^5$, then there is a pencil
of cubics passing through $X$, {\it i.e.} we have a linear
family $\{ D_{t \in {\bf P}^1} \}$ of cubic fourfolds, so that
$X = \cap_{t \in {\bf P}^1} D_t$.  The object corresponding to $X$ 
through taking linear sections in noncommutative HPD will be
a noncommutative Calabi-Yau three-fold $Y$ which is naturally fibered
over the ${\bf P}^1$ that parametrizes the pencil.
We can see this by constructing $Y$, as follows.
Start with the universal family
\begin{displaymath}
U \: = \: \{ (x,t) \in {\bf P}(V) \times {\bf P}^1 |
x \in D_t \}
\end{displaymath}
of fourfolds in the pencil.  Let $p: U \rightarrow
{\bf P}^1$ be the natural projection.  Consider the full subcategory
${\cal A}_X \subset D^b(U)$, defined as the left orthogonal
\begin{displaymath}
{}^{\perp}\left\langle p^* D^b({\bf P}^1) \otimes {\cal O}(-3),
p^* D^b({\bf P}^1) \otimes {\cal O}(-2),
p^* D^b({\bf P}^1) \otimes {\cal O}(-1) \right\rangle
\end{displaymath}
Again, it is straightforward to check that if $X$ is smooth then
${\cal A}_X$ is a smooth compact three-dimensional Calabi-Yau category.

We will think of ${\cal A}_X$ as the derived category of a noncommutative
K3 $Y$ which is fibered over ${\bf P}^1$ by noncommutative K3 surfaces
(defined as in (2)).  If we specialize $X$ to be the complete intersection
of two Pfaffian cubics, then $Y$ will be an ordinary Calabi-Yau which is
fibered by K3 surfaces of degree 14.
In fact, in his paper on the Grassmannian-Pfaffian HPD \cite{kuz3} Kuznetsov
shows that in case $Y$ is commutative we have an equivalence of
categories $D^b(X) \cong D^b(Y)$.

Thus, homological projective duality can be applied to more
examples than just, complete intersections of quadrics
(or hyperplanes in Grassmannians); it makes predictions for
other cases, as we have seen in this example, that correctly
match physics.

Unfortunately, it is very difficult to check whether analogous statements
are true for GLSM's describing hypersurfaces of degree
greater than three in projective
spaces.  We conjecture that homological projective duality applies in
general, but will have to leave a complete verification to future work.

\section{Conclusions}

In this paper, we have done a number of different things.
By studying the properties of some basic, ordinary-seeming gauged
linear sigma models such as that for the complete intersection
${\bf P}^7[2,2,2,2]$, we have discovered in many examples
that the Landau-Ginzburg point
of the GLSM realizes geometry in a novel fashion, not as a complete
intersection; that same geometry is not birational to the large-radius
geometry, providing a GLSM interpolating between two non-birational
geometries.  We have provided a mathematical interpretation
(homological projective duality) to substitute for `birational'
in relating GLSM phases, an interpretation which often even
covers non-geometric Landau-Ginzburg models.
We have discovered that many of those Landau-Ginzburg points do
not flow under the renormalization group to nonlinear sigma models,
or any other known type of conformal field theory,
but rather give a physical realization of certain `noncommutative
resolutions' of spaces, and so give us a way of defining conformal
field theories for nc spaces (in Kontsevich's sense), a whole new
kind of CFT and string compactification.
In the analysis of the Landau-Ginzburg points of these otherwise
ordinary-seeming GLSM's, we have also found that understanding stacks
and how stacks appear in physics plays a crucial role, as the analysis
of the Landau-Ginzburg points relies crucially on facts concerning the
behavior of strings propagating on gerbes.
Thus, stacks play a much more important role in the physical understanding
of many GLSM's than previously realized.

One of the open questions in this paper is, to what extent can all
phases of gauged linear sigma models be understood via homological projective
duality?  To answer this question, the mathematics of homological projective
duality needs to be much better understood.
For example, it is known that flops of small resolutions of
the most basic conifold singularity
in three complex dimensions are examples of homological projective
duality, but whether more general flops are also related by homological
projective duality, is not yet known.

In the same vein, one can ask, can the McKay correspondence or any
of its variants be understood as forms of homological projective duality,
especially since various forms of McKay can often be realized physically
as different phases of gauged linear sigma models.

Another set of open problems arises from the fact that we have discovered
more conformal field theories, not of a form previously discussed
-- in fact, we have even found a description of them with gauged linear
sigma models.  The `noncommutative resolutions' arising at Landau-Ginzburg
points are, close to geometric singularities, not even locally spaces.
These noncommutative resolutions do not obviously have any description
as any sort of conformal field theory discussed elsewhere in the literature.
As such, the properties of these new conformal field theories needs to
be studied.  We have briefly described some properties of the B-branes
in these conformal field theories, but nothing else.
For example, it would be very interesting to compute Gromov-Witten
invariants of these theories.
(Work on other Gromov-Witten variations that might be applicable
here will be discussed in \cite{mejosh}.)

\section{Acknowledgements}

We would like to thank R.~Donagi, M.~Gross,
A.~Kuznetsov, and I.~Melnikov for useful conversations.
In particular, we would especially like to thank
A.~Kuznetsov for many patient explanations, valuable discussions,
and for sharing his insights.
Parts of this work were done during the Aspen Center for Physics
program ``String Theory and Quantum Geometry,'' July 2007.
J.D. and T.P. would like to thank the ACP for providing excellent
working conditions and a stimulating research environment.
A.C. was partially supported by NSF grant DMS-0556042.
J.D. was partially supported by NSF grant PHY-0455649.
S.H. was partially supported by DOE grant DE-FG02-90ER40542.
T.P. was partially supported by NSF Focused Research Grant
DMS-0139799, NSF Research Training Group Grant DMS-0636606,
and NSF grant DMS-0700446.
E.S. was partially supported by NSF grant DMS-0705381.
S. H. is the D. E. Shaw \& Co., L. P. Member at the Institute for Advanced
Study.

\appendix
 
\section{Calabi-Yau categories and noncommutative spaces}
\label{cy-categories}

\subsection{Definitions}

The notion of a smooth and proper Calabi-Yau category was introduced
by Kontsevich in 1998 \cite{kontsevich1} and was subsequently discussed
in detail in
\cite{kontsevich2,soibelman,costello1,tv1}.  
It abstracts the notion of a smooth and compact Calabi-Yau background that
can be used as a target in (A or B twisted) non-linear sigma models.
The basic idea is to replace the Calabi-Yau space by the category of
A or B branes.

Consider a small pre-triangulated\footnote{
The derived category is then the homotopy category of this
pre-triangulated category.
} Karoubi 
closed\footnote{Karoubi closed means that every projector splits.
Recall that a projector on an object $E$ in a category is an
endomorphism $\pi: E \rightarrow E$ such that $\pi \circ \pi = \pi$.
We say a projector $\pi$ splits if there is a direct sum decomposition
$E = E_1 \oplus E_2$ such that $\pi$ is the projection map
$E \rightarrow E_1$.  Then, a category is called Karoubi closed if 
any projector on any object splits.}
dg-category ${\cal C}$.  
We think of the objects of this category as complexes of sheaves on
a `noncommutative space' or `nc space' $X$, where $X$ is defined by
the category ${\cal C}$.
If we were trying to model an actual complex manifold $X$, then we would
take ${\cal C}$ to be the (dg-enhancement of) the derived category of
coherent sheaves on $X$.  For that matter, $X$ might turn out to
be a Calabi-Yau with a flat $B$ field, or even a Landau-Ginzburg model,
by virtue of matrix factorizations -- $X$ need not be an actual space.
`Noncommutative'
is a misnomer in this context, as there need not be a noncommutative algebraic
structure associated with $X$.

An nc space $X$ is called {\it algebraic} if its sheaf theory is computed by 
an algebra, {\it i.e.} if we can find a dg algebra $A = (A^{\cdot}, d)$
over ${\bf C}$, such that the category $C_X$ is equivalent to the dg-category
of (perfect) dg-modules over $A$.

An nc space $X$ is called {\it smooth} if it is algebraic and if the dg
algebra $A$ that computes its sheaf theory is perfect when viewed as a 
bimodule over itself.

An nc space $X$ is called {\it compact} if it is algebraic and the differential
on the dg algebra $A$ has finitely many, finite dimensional cohomology
groups.  That is, $\mbox{dim}_{{\bf C}} H^{\cdot}(A^{\cdot}, d) < \infty$.

It is known that:
\begin{itemize}
\item The dg category of complexes of injective sheaves with
coherent cohomology on any complex algebraic variety is
computed by a dg algebra.  Thus, every complex variety is algebraic
when viewed as an nc space.
Also, non-algebraic complex manifolds may not be algebraic nc spaces.
For example, a generic non-algebraic K3 is not an algebraic nc space.
One can define sheaves on non-algebraic K3's, so there is a category,
and hence an nc space; however, by a theorem of Bondal and van den Bergh
\cite{bondaletal},
the category of sheaves on a generic non-algebraic K3 is not saturated,
which means there is no smooth and proper algebra that computes this category.
Hence non-algebraic K3's are nc spaces, but not algebraic nc spaces.

\item A complex variety $X$ is proper or smooth if and only if it is
proper or smooth when we view it as a nc space.
\end{itemize}

There are many other examples of proper and smooth nc spaces coming
from stacks, deformation quantizations, symplectic manifolds, generalized
complex manifolds, and Landau-Ginzburg models.

If $X$ is a proper and smooth nc space, then one can show that the category
$C_X$ defining $X$ has a Serre functor $S: C_X \rightarrow C_X$.

We say that a proper smooth nc space $X$ is an nc Calabi-Yau, if the
Serre functor is a power of the shift functor, {\it i.e.}, if
$S = [d]$ for some positive integer $d$.  (The category defining an
nc Calabi-Yau is called a Calabi-Yau category.)
The number $d$ is the dimension
of the nc Calabi-Yau $X$.  Notice that one can define nc Calabi-Yau spaces
of fractional dimension $p/q$ by requiring that $S^q = [p]$.  
One example is furnished by Enriques surfaces,
where the square of the Serre functor is $[2]$, and hence the
(fractional) dimension is $2/2 = 1$.
(In particular, note the notion of dimension can be slightly misleading,
as it leads to an Enriques surface having dimension 1 as an nc space
but dimension 2 as an ordinary complex manifold.)
There are many more examples of such spaces and one should find their
meaning eventually.

\subsection{Deformations}

If we start with an ordinary Calabi-Yau space $X$, we can look at all
deformations of $X$ viewed as an nc space.  In other words,
we can forget about the actual $X$ but deform the category $C_X$ as a
dg category.  The tangent space to such deformations can be identified with
the second Hochschild cohomology of $X$.  In the smooth proper case this
tangent space has a decomposition
\begin{displaymath}
HH^2(X) \: = \: H^2(X, {\cal O}_X) 
\oplus H^1(X, T_X) \oplus H^0(X, \Lambda^2 T_X)
\end{displaymath}
where each piece can be interpreted geometrically:
\begin{itemize}
\item $H^2(X, {\cal O}_X)$ are the deformations of $X$ as an ${\cal O}^{\times}$
gerbe;
\item $H^1(X, T_X)$ are the usual deformations of $X$ as a variety;
\item $H^0(X, \Lambda^2 T_X)$ are the deformation quantizations of the
product structure on the structure sheaf of $X$.
\end{itemize}

In the example in section~\ref{hpdnck3} we start with $S$ which is a 
K3 surface of degree 14, and corresponds to a Pfaffian cubic fourfold
$D \subset {\bf P}^5$ under the Grassmannian-Pfaffian duality.
Next we look at the nc deformations of $S$ that correspond to deforming
the fourfold $D$ to a non-Pfaffian cubic.  In particular, if
$\xi$ is a deformation direction for $D$,
then $\xi$ can be interpreted as an element in $HH^2(S)$ and hence breaks 
into three components $\xi = \xi_{20} + \xi_{11} + \xi_{02}$.
If $\xi$ moves $D$ to a non-Pfaffian cubic, then we can not have
$\xi = \xi_{11}$.  
In the example in section~\ref{hpdnck3},
an old geometric argument with varieties of lines shows that the gerby part
$\xi_{20}$ of the deformation direction $\xi$ is non-zero,
and it is natural to expect that $\xi_{02}$ is also non-zero and so
this nc deformation of $S$ of of
`mixed' nature.

\subsection{Cohomology of nc spaces}

Suppose that $X$ is an algebraic nc space and let $A$ be a dg algebra
that computes the sheaf theory of $X$.  
(Recall that a nc space $X$ is defined by a category $C$, but for $X$
to be an algebraic nc space means that the category $C$ is the
category of perfect dg-modules over some dg algebra; $A$ is that dg algebra.)

Then we can attach two natural
cohomology theories to $X$:
\begin{itemize}
\item The Dolbeault cohomology $H^{\cdot}_{Dol}(X)$ of $X$.  In terms of
$A$ we have $H^{\cdot}_{Dol} \equiv HH_{\cdot}(A)$, {\it i.e.}
$H^{\cdot}_{Dol}(X)$ is the Hochschild {\it homology} of the dg algebra
$A$.
\item The de Rham cohomology $H^{\cdot}_{dR}(X)$ of $X$.  In terms of 
the dg algebra $A$ we have $H^{\cdot}_{dR}(X) \equiv HP_{\cdot}(A)$,
{\it i.e.} $H^{\cdot}_{dR}(X)$ is the {\it periodic cyclic homology}
of the dg algebra $A$.
\end{itemize}

These definitions are justified by the observation that when $X$ is a smooth
variety we have
\begin{itemize}
\item (Kostant-Hochschild-Rosenberg) $HH_k(A) = \oplus_{p-q=k} H^p(X, 
\Omega^q_X)$
\item (Weibel) $HP_{even/odd}(A) = H^{even/odd}_{dR}(X)$.
\end{itemize}
If $S$ is a nc K3 surface defined as above from a general cubic fourfold
$D$, then $H^{\cdot}_{Dol}(S)$ is a subspace in the ordinary
Dolbeault cohomology $H^{\cdot}_{Dol}(D)$ which is the orthogonal
complement to $\{ \mbox{ch}({\cal O}(-i)) \}_{i=1,2,3}$ with respect to the
Mukai pairing on $H^{\cdot}_{Dol}(D)$.


\begin{thebibliography}{199}

\addcontentsline{toc}{section}{References}


\bibitem{WitPhases} E. Witten, ``Phases of N=2 theories in
two dimensions,''
Nucl Phys. {\bf B403} (1993) 159-222, {\tt hep-th/9301042}.

\bibitem{ps4} S. Hellerman, A. Henriques, T. Pantev, E. Sharpe,
M. Ando, ``Cluster decomposition, T-duality, and gerby CFT's,''
{\tt hep-th/0606034}.

\bibitem{hori1} K. Hori, D. Tong, ``Aspects of non-Abelian gauge
dynamics in two-dimensional $N=(2,2)$ theories,''
{\tt hep-th/0609032}.

\bibitem{estx} E. Sharpe, ``String orbifolds and quotient stacks,''
Nucl. Phys. {\bf B627} (2002) 445-505,
{\tt hep-th/0102211}.

\bibitem{vafawitten} C. Vafa and E. Witten, ``On orbifolds with
discrete torsion,'' J. Geom. Phys. {\bf 15} (1995) 189-214,
{\tt hep-th/9409188}.

\bibitem{meron} R. Donagi, E. Sharpe, ``GLSMs for partial flag manifolds,''
{\tt arXiv:  0704.1761}.

\bibitem{reidthesis} M. Reid, ``The complete intersection of two or
more quadrics,'' Ph.D. thesis, Trinity College, Cambridge, 1972,
available at {\tt http://www.maths.warwick.ac.uk/\v{~}miles/}

\bibitem{gh} P. Griffiths and J. Harris, {\it Principles of Algebraic
Geometry}, John Wiley \& Sons, New York, 1978.

\bibitem{clemens1} H. Clemens, ``Double solids,''
Adv. in Math. {\bf 47} (1983) 107-230.

\bibitem{cynk1} S. Cynk and C. Meyer, ``Geometry and arithmetic
of certain double octic Calabi-Yau manifolds,''
{\tt math.AG/0304121}.

\bibitem{meshelandrei} A. C\u ald\u araru, S. Katz, E. Sharpe,
``D-branes, $B$ fields, and Ext groups,''
Adv. Theor. Math. Phys. {\bf 7} (2004) 381-404,
{\tt hep-th/0302099}.

\bibitem{gp} M. Gross, S. Pavanelli, ``A Calabi-Yau threefold with
Brauer group $({\bf Z}_8)^2$,'' {\tt math.AG/0512182}.

\bibitem{grosspriv} M. Gross, private communication, September 27, 2006.

\bibitem{kuz2} A. Kuznetsov, ``Derived categories of quadric fibrations and
intersections of quadrics,'' {\tt math.AG/0510670}.

\bibitem{medc} E. Sharpe, ``D-branes, derived categories,
and Grothendieck groups,'' Nucl. Phys. {\bf B561} (1999) 433-450,
{\tt hep-th/9902116}.

\bibitem{dougdc} M. Douglas, ``D-branes, categories, and
$N=1$ supersymmetry,'' J. Math. Phys. {\bf 42} (2001) 2818-2843,
{\tt hep-th/0011017}.

\bibitem{medcrev} E. Sharpe, ``Lectures on D-branes and sheaves,''
writeup of lectures given at the twelfth Oporto meeting on ``Geometry,
Topology, and Physics,'' and at the Adelaide workshop
``Strings and Mathematics 2003,'' {\tt hep-th/0307245}.

\bibitem{ps1} T. Pantev, E. Sharpe, ``Notes on gauging noneffective
group actions,'' {\tt hep-th/0502027}.

\bibitem{ps2} T. Pantev, E. Sharpe, ``String compactifications on
Calabi-Yau stacks,'' Nucl. Phys. {\bf B733} (2006) 233-296,
{\tt hep-th/0502044}.

\bibitem{stxglsm} T. Pantev, E. Sharpe, ``GLSM's for gerbes (and other
toric stacks),'' Adv. Theor. Math. Phys. {\bf 10} (2006) 77-121,
{\tt hep-th/0502053}.

\bibitem{ps5} E. Sharpe, ``Derived categories and stacks in
physics,'' {\tt hep-th/0608056}.

\bibitem{kuz1} A. Kuznetsov, ``Homological projective duality,''
{\tt math.AG/0507292}.

\bibitem{kuz3} A. Kuznetsov, ``Homological projective duality for
Grassmannians of lines,'' {\tt math.AG/0610957}.

\bibitem{harris} J. Harris, {\it Algebraic geometry:  a first course},
Grad. texts in math. 133, Springer-Verlag, New York, 1992.

\bibitem{andrei-ell} A. C\u ald\u aru, ``Derived categories of twisted
sheaves on elliptic threefolds,'' {\tt math.AG/0012083}.

\bibitem{andrei-nicadd} A. C\u ald\u aru, Nicholas Addington, work in
progress.

\bibitem{kapli1} A. Kapustin, Y. Li, ``D-branes in Landau-Ginzburg models and
algebraic geometry,'' {\tt hep-th/0210296}.

\bibitem{kuzpriv} A. Kuznetsov, private communication.

\bibitem{mukai1} S. Mukai, ``Moduli of vector bundles on K3 surfaces,
and symplectic manifolds,'' Sugaku Expositions {\bf 1} (1988) 139-174.

\bibitem{kuzpriv129} A. Kuznetsov, private communication,
January 29, 2007.

\bibitem{andreitoappear} A. C\u ald\u araru, to appear.

\bibitem{dt3} E. Sharpe, ``Discrete torsion,'' Phys. Rev. {\bf D68} (2003)
126003, {\tt hep-th/0008154}. 

\bibitem{dtrev} E. Sharpe, ``Recent developments in discrete torsion,''
Phys. Lett. {\bf B498} (2001) 104-110,
{\tt hep-th/0008191}.

\bibitem{agw} A. C\u ald\u araru, A. Giaquinto, S. Witherspoon,
``Algebraic deformations arising from orbifolds with discrete
torsion,'' {\tt math.AG/0210027}.

\bibitem{ilarionronen} I. Melnikov, R. Plesser, ``A-model correlators
from the Coulomb branch,'' JHEP {\bf 0602} (2006) 044,
{\tt hep-th/0507187}.

\bibitem{bertrampriv} A. Bertram, private communication,
January 5, 2007.

\bibitem{beauville} A. Beauville, {\it Complex algebraic surfaces},
second edition, Cambridge University Press, 1996.

\bibitem{simpson} J. Iyer, C. Simpson, ``A relation between the parabolic
Chern characters of the de Rham bundles,''
{\tt math.AG/0603677}.

\bibitem{evaallan} A. Adams, J. Polchinski, E. Silverstein,
``Don't panic!  Closed string tachyons in ALE space-times,''
JHEP {\bf 0110} (2001) 029, {\tt hep-th/0108075}.

\bibitem{greg1} J. Harvey, D. Kutasov, E. Martinec, G. Moore,
``Localized tachyons and RG flows,''
{\tt hep-th/0111154}.

\bibitem{greg2} E. Martinec, G. Moore, ``On decay of K theory,''
{\tt hep-th/0212059}.

\bibitem{kuzpriv310} A. Kuznetsov, private communication,
March 10, 2007.

\bibitem{orlov1} D. Orlov, ``Triangulated categories of singularities
and D-branes in Landau-Ginzburg models,''
{\tt math.AG/0302304}.

\bibitem{orlov2} D. Orlov, ``Triangulated categories of singularities
and equivalences between Landau-Ginzburg models,''
{\tt math.AG/0503630}.

\bibitem{orlov3} D. Orlov, ``Derived categories of coherent sheaves and 
triangulated categories of singularities,''
{\tt math.AG/0503632}.


\bibitem{metonyinprogress} T. Pantev, E. Sharpe, work in progress.

\bibitem{mejosh} J. Guffin, E. Sharpe, to appear.

\bibitem{kontsevich1}  M. Kontsevich, ``Course on non-commutative
geometry,'' ENS, 1998, lecture notes at
{\tt http://www.math.uchicago.edu/\v~arinkin/langlands/kontsevich.ps}.

\bibitem{kontsevich2} M. Kontsevich, talk at the ``Hodge centennial
conference,'' Edinburgh, 2003.

\bibitem{soibelman} Y. Soibelman, ``Lectures on deformation theory and
mirror symmetry,'' IPAM, 2003,
{\tt http:// www.math.ksu.edu/\v~soibel/ipam-final.ps}.

\bibitem{costello1} K. Costello, ``Topological conformal field
theories and Calabi-Yau categories,'' {\tt math.QA/0412149}.

\bibitem{tv1} B. To\"en, M. Vaquie, ``Moduli of objects in dg-categories,''
{\tt math.AG/0503269}.

\bibitem{bondaletal} A. Bondal, M. van den Bergh, ``Generators and 
representability of functors in commutative and noncommutative
geometry,'' Mosc. Math. J. {\bf 3} (2003) 1-36, 258.


















\end{thebibliography}
\end{document}